\documentclass[twocolumn,english,prl,showpacs,preprintnumbers,superscriptaddress]{revtex4-2}
\usepackage[utf8]{inputenc}
\usepackage{color}
\usepackage{amsmath,amssymb}
\usepackage{amssymb}
\usepackage{graphicx}
\usepackage{textcomp}
\usepackage{dcolumn}
\usepackage{bm}
\usepackage[right]{eurosym}
\usepackage{float}
\usepackage[english]{babel}
\usepackage{blindtext}
\usepackage{babel}
\usepackage[normalem]{ulem}
\usepackage{wasysym}
\usepackage{hyperref}
\pdfoutput=1
\mathchardef\mhyphen="2D

\begin{document}
\title{Interatomic Coulombic decay in lithium-doped large helium nanodroplets induced by photoelectron impact excitation}

\author{L. Ben Ltaief}
\affiliation{Department of Physics and Astronomy, Aarhus University, 8000 Aarhus C, Denmark}

\author{K. Sishodia}
\affiliation{Quantum Center of Excellence for Diamond and Emergent Materials and Department of Physics, Indian Institute of Technology Madras, Chennai 600036, India}

\author{J. D. Asmussen}
\affiliation{Department of Physics and Astronomy, Aarhus University, 8000 Aarhus C, Denmark}

\author{A. R. Abid}
\affiliation{Department of Physics and Astronomy, Aarhus University, 8000 Aarhus C, Denmark}

\author{S. R. Krishnan}
\affiliation{Quantum Center of Excellence for Diamond and Emergent Materials and Department of Physics, Indian Institute of Technology Madras, Chennai 600036, India}

\author{H. B. Pedersen}
\affiliation{Department of Physics and Astronomy, Aarhus University, 8000 Aarhus C, Denmark}

\author{N. Sisourat}
\affiliation{Sorbonne Université, CNRS, Laboratoire de Chimie Physique Matière et Rayonnement, UMR 7614, F-75005 Paris, France}

\author{M. Mudrich}
\affiliation{Department of Physics and Astronomy, Aarhus University, 8000 Aarhus C, Denmark}

\date{November 2024}

\begin{abstract}
Irradiation of condensed matter with ionizing radiation generally causes direct photoionization as well as secondary processes that often dominate the ionization dynamics. Here, large helium (He) nanodroplets with radius $\gtrsim40~$nm doped with lithium (Li) atoms are irradiated with extreme ultraviolet (XUV) photons of energy $h\nu\geq 44.4$~eV and indirect ionization of the Li dopants is observed in addition to direct photoionization of the He droplets. Specifically, Li ions are efficiently produced by an interatomic Coulombic decay (ICD) process involving  metastable He$^*$ atoms and He$_2^*$ excimers which are populated by elastic and inelastic scattering of photoelectrons in the nanodroplets as well as by electron-ion recombination. This type of indirect ICD, observed in large He nanodroplets in nearly the entire XUV range, turns out to be more efficient than Li dopant ionization by ICD following direct resonant photoexcitation at $h\nu=21.6~$eV and by charge-transfer ionization. Indirect ICD processes induced by scattering of photoelectrons likely play an important role in other condensed phase systems exposed to ionizing radiation as well, including biological matter. 
\end{abstract}

\maketitle

\section{Introduction}
Interaction of extreme ultraviolet (XUV) or x-ray radiation with condensed-phase systems generally causes primary ionization or excitation which is often followed by \textit{intra}-atomic/molecular decay processes, by secondary processes due to electron scattering, or by \textit{inter}-atomic/molecular transfer of energy or charge within the medium. In a weakly bound system, \textit{e.~g.} a van-der-Waals or hydrogen-bonded cluster, the energy deposited in the electronic system by photon absorption can be released by \textit{intra}-atomic/molecular processes such as fluorescence emission or Auger-Meitner decay. Alternatively, \textit{inter}atomic/molecular decay processes involving neighboring sites can occur, such as interatomic/molecular Coulombic decay (ICD) in which energy is transferred from one excited atom or molecule to a neighboring one which in turn is ionized~\cite{Jahnke:2020}. ICD has captured considerable interest since its discovery~\cite{Cederbaum:1997} and has been characterized in detail for various small van-der-Waals clusters~\cite{Jahnke:2020, Hergenhahn:2011}. More recently, the focus has shifted to ICD in more extended condensed-phase systems, in particular aqueous solutions, due to their relevance for radiobiology~\cite{Stumpf:2016, Ren:2018, Zhang:2022, Ren:2023, gopakumar2023radiation}. 

He nanodroplets are a special type of condensed-phase model system; atoms and molecules attached to He nanodroplets are highly mobile due to their quantum fluid properties, and electronically excited species tend to form void bubbles which are expelled to the droplet surface~\cite{Mudrich:2020}. Owing to their large pick-up cross section, He nanodroplets are readily doped with foreign atoms and molecules to study high-resolution molecular spectra and guest-host interactions in this fundamentally important system~\cite{Toennies:2004, Mudrich:2014,slenczka2022molecules}. 

Both pure and doped He nanodroplets have previously proven well suited as a model system for elucidating ICD and related processes using weak synchrotron radiation~\cite{Froechtenicht:1996,BuchtaJCP:2013, LaForgePRL:2016, Shcherbinin:2017, Shcherbinin:2018, wiegandt:2019, Ltaief:2019, Ltaief:2020, ltaief:2023, asmussen2023dopant, asmussen2023secondary, ltaiefPRR:2024,bastian2024observation}, XUV pulses from high harmonic generation~\cite{jurkovivcova:2024} or intense XUV pulses from a free electron laser (FEL)~\cite{Ovcharenko:2014, LaForge:2014, Ovcharenko:2020, LaForge:2021, Asmussen:2021, michiels:2021,laforge2022relaxation}. In those studies, He droplets were excited either to their 1s2p or 1s3p/4p-correlated absorption bands~\cite{Froechtenicht:1996, Buchta:2013, Ltaief:2019, Shcherbinin:2018, asmussen2023dopant, asmussen2023secondary, ltaiefPRR:2024, jurkovivcova:2024}, or to high-lying or ionized states~\cite{LaForgePRL:2016, Shcherbinin:2017, wiegandt:2019, Ltaief:2020, ltaief:2023, bastian2024observation} and either He or dopant atoms or molecules were ionized by ICD processes.

An important aspect of ICD is that it often produces low-energy electrons (LEE) which are proven to efficiently cause radiation damage in biological matter~\cite{Boudaiffa:2000, Alizadeh:2015}. In extended weakly bound systems, however, slow secondary electrons can also be produced by inelastic scattering of fast photoelectrons or Auger electrons~\cite{pimblott2007} and may mask the LEE created by ICD~\cite{barth2005, mucke2015, Iablonskyi:2016, ltaief:2018, malerz2021low}. Therefore, it is often hard to clearly identify ICD in condensed-phase systems as scattered electrons tend to obscure its signature in electron spectra. 

On the other hand, electron scattering can play an important role in initiating ICD processes in extended systems, as we have recently demonstrated~\cite{ltaief:2023, ltaiefPRR:2024}. In large pure He nanodroplets, two or more excited He$^*$ atoms can be formed in one droplet upon irradation with energetic XUV photons; the photoelectron emitted from a He atom in a droplet subsequently scatters on neighboring He atoms and is then recaptured by the parent He$^+$ ion. While the first He$^*$ atom is formed by an inelastic electron-He collision inducing impact excitation, the second one is formed by electron-He elastic scattering causing the electron to slow down until it recombines with the parent He ion. The resulting pair of excited He$^*$ atoms can in turn electronically relax by ICD leading to the emission of characteristic electrons.
This ICD was found to be the dominant electron emission process in large pure He nanodroplets with radius $\gtrsim40~$nm in nearly the entire range of XUV photon energies from the photoelectron impact excitation threshold of He ($h\nu=44.4~$eV) up to the soft x-ray range~\cite{ltaief:2023}. It likely plays an important role in other dense systems exposed to ionization radiation as well, including biological matter. 

So far, ICD induced by photoelectron scattering has been studied only for large pure He nanodroplets~\cite{ltaief:2023, ltaiefPRR:2024}. Here, we present a systematic study of a similar ICD process induced by photoelectron impact excitation of large He nanodroplets doped with Li atoms. Owing to the simple electronic structure of the He atom and the peculiar quantum-fluid properties of He nanodroplets~\cite{Toennies:2001}, we measure highly resolved electron spectra for various secondary ionization channels of Li dopants involving ICD of impact-excited He nanodroplets. The comparison with spectra measured for resonantly photoexcited doped He droplets at $h\nu=21.6~$eV reveals characteristic differences due to different initial states of the impact-excited respectively photoexcited He$^*$. Our results are supported by \textit{ab initio} calculations of ICD rates involving He$^*$ in the metastable 1s2s\,$^1$S singlet and the 1s2s\,$^3$S triplet states.

\begin{figure}[t!]
	\center
    \includegraphics[width=0.65\columnwidth]{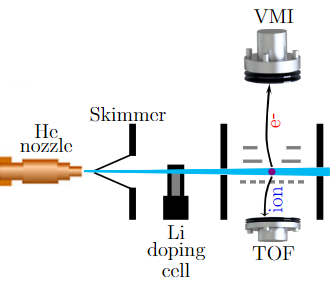}\caption{\label{fig0} Sketch of the experimental setup used in this work. It consists of a He nanodroplet beam source, a heated cell for doping the droplets with lithium atoms and a combined ion time-of-flight (TOF) and electron velocity-map imaging (VMI) spectrometer. The photon beam intersects the He nanodroplet beam at right angles in the center of the spectrometer.}
\end{figure}
\section{Experimental methods}
The experiment was performed at the XENIA endstation of the AMO beamline at the synchrotron radiation facility ASTRID2 Aarhus, Denmark~\cite{bastian2022}. The setup is schematically depicted in Fig.~1 and has been described in detail elsewhere~\cite{bastian2022}. In short, a continuous beam of He nanodroplets containing $3\times10^3$ up to $\sim 4\times10^7$ He atoms per droplet is generated by expanding He out of a cryogenic nozzle at a temperature ranging from 18 down to 7~K and a He backing pressure of 30~bar. The He nanodroplets are doped with Li atoms by passing the droplets through a radiatively heated vapor cell of length 10~mm. The temperature of the heated Li doping cell was varied in the range of $T_\mathrm{Li}= 190$-$410\,^\circ$C. At $T_\mathrm{Li}= 380\,^\circ$C we estimate the number of Li atoms picked up by the droplets with radius $R=75~$nm to about 1200 Li atoms using the simple model reported in~\cite{kuma2007laser}. The exact number of Li atoms remaining bound to the droplet surface after coagulating into clusters is hard to estimate as it depends on the size and spin state of the formed Li clusters~\cite{Buenermann:2011}.
A mechanical beam chopper is placed between the skimmer and the doping cell for discriminating all the shown spectra from the background.

In the detector chamber further downstream, the He droplet beam intersects the XUV photon beam at right angles in the center of a combined electron velocity-map imaging (VMI) and ion time-of-flight (TOF) spectrometer operated in electron-ion coincidence detection mode. The photon flux is $\Phi\approx 2.1\times10^{14}$~photons~s$^{-1}$cm$^{-2}$ at $h\nu=46.2~$eV and $\Phi\approx 1.3\times10^{14}$~s$^{-1}$cm$^{-2}$ at $h\nu=21.6~$eV, respectively. Electron spectra are inferred from electron images by Abel inversion~\cite{Dick:2014}. This setup allows us to simultaneously measure spectra and angular distributions of all emitted electrons and of electrons emitted in coincidence with specific fragment ions in a wide range of photon energies.

\section{Results and discussion}
\subsection{Ion mass spectra}
\begin{figure}[t!]
	\center
    \includegraphics[width=0.95\columnwidth]{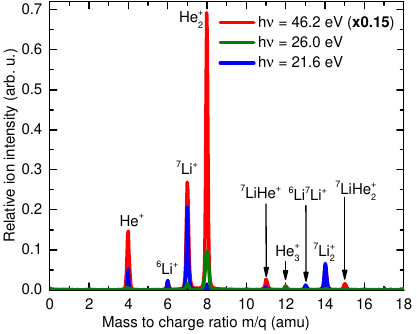}\caption{\label{fig:massspec} a) Mass spectra measured for large He nanodroplets ($R=50$~nm) doped with Li atoms at the photon energies $h\nu=21.6~$eV (blue line), $h\nu=26.0~$eV (green line) and $h\nu=46.2~$eV (red line). All spectra are normalized to the photon flux and to the $h\nu$-dependent absorption cross section of He.}
\end{figure}

We start the presentation of the experimental data by comparing mass spectra recorded by irradiating large He nanodroplets at various characteristic photon energies. At $h\nu=21.6~$eV (blue line in Fig.~\ref{fig:massspec}), He droplets are resonantly photoexcited into their strongest absorption band correlating to the 1s2p state of He~\cite{Joppien:1993,Buchta:2013}. For small droplets with radius $R\lesssim 10$~nm, only dopant ions, Li$^+$, Li$_2^+$ and to a lesser extent [LiHe]$^+$ and [LiHe]$_2^+$, are expected due to ICD of Li dopants by interaction with photoexcited He$^*$. This process is traditionally termed Penning ionization~\cite{Scheidemann:1997,Buchta:2013,Ltaief:2019}. In large He droplets with radii $R\gtrsim 10$~nm containing $N\gtrsim 10^5$ He atoms, a small fraction of the droplets are multiply excited even at the low intensities available at synchrotrons owing to the droplets' large total absorption cross section $N\times\sigma\gtrsim 10^5 \times 25$~Mb, where $\sigma\approx 25$~Mb is the estimated absorption cross section of a He atom in a nanodroplet~\cite{BuchtaJCP:2013,Ovcharenko:2014,Ovcharenko:2020}. The long lifetime $\gtrsim 1.6~\mu$s~\cite{McKinsey:2003} of the metastable He$^*$'s allows the droplets to accumulate He$^*$'s over many consecutive synchrotron pulses~\cite{BuchtaJCP:2013}. Pairs of He$^*$'s attached to one droplet efficiently autoionize by ICD, thereby producing mainly He$^+$, He$_2^+$, and to a lesser extent He$_3^+$ ions~\cite{ltaiefPRR:2024}. The fact that the Li$^+$ ion yield exceeds that of He$_n^+$ for large He droplets despite the small dopant-to-He ratio of $\lesssim10^{-3}$ reflects the low probability ($\lesssim10^{-3}$) of droplets being multiply excited even at the strong 1s2p resonance~\cite{ltaiefPRR:2024}.

At $h\nu=26~$eV (green line in Fig.~\ref{fig:massspec}), \textit{i.~e.} just above the ionization potential of He, $E_i = 24.6$~eV, He ions He$_n^+$, $n=1,\,2,\,3$, are created by direct photoionization and Li$^+$ and Li$_2^+$ ions are produced by electron transfer from the Li dopant to the He ion~\cite{Buchta:2013}. As compared to resonant ICD, electron transfer leading to free Li$^+$ and Li$_2^+$ ions is less efficient~\cite{asmussen2023dopant}. This has been rationalized by the tendency of He$^+$ ions to solvate in the droplet interior by forming stable complexes~\cite{haberland1995absorption}, whereas Li dopants are located at the droplet surface just like He$^*$ and He$_2^*$ excitations after relaxation of the droplet~\cite{Scheidemann:1997,Buchta:2013,Ltaief:2019,Mudrich:2020}.  

The most important new observation reported here is that at $h\nu=46.2~$eV (red line in Fig.~\ref{fig:massspec}), the yield of both He and Li ions exceeds that at $h\nu=21.6$ and $26~$eV by more than one order of magnitude when taking into account the photon energy-dependent He absorption cross section~\cite{Samson:2002}. Thus, for every photon absorbed by a He nanodroplets we find that the ejection of a He ion or a Li ion is much more efficient at $h\nu=46.2~$eV, where photoelectrons have a high cross section for exciting He$^*$ atoms by inelastic scattering. Additionally, a slow electron emerging from an inelastic scattering event around the threshold for He$^*$ impact excitation, \textit{i.~e.} at $h\nu\approx 45$-$50$~eV, can recombine with its parent ion to form a second He$^*$ in the same droplet~\cite{ltaief:2023}. Pairs of metastable He$^*$'s formed in this way subsequently decay by the reaction
\begin{equation}
\label{eq:He-ICD}
\mathrm{He}^* + \mathrm{He}^* \rightarrow \mathrm{He} + \mathrm{He}^+ + e_\mathrm{ICD}
\end{equation} 
as we have previously studied~\cite{ltaief:2023}. This process manifests itself as a pronounced peak in electron spectra at the characteristic energy $E_e^\mathrm{ICD}\approx 15$~eV of the ICD electron, $e_\mathrm{ICD}$. The observation of high yields of Li ions under these conditions suggests that the analogous ICD reaction of Li dopants in the regime of electron inelastic scattering,
\begin{equation}
\label{eq:Li-ICD}
    \mathrm{He}^* + \mathrm{Li} \rightarrow \mathrm{He} + \mathrm{Li}^+ + e_\mathrm{ICD},
\end{equation} 
is highly efficient as well. In particular, it seems to be by far more efficient in producing free Li ions from large He droplets than electron-transfer ionization, $\mathrm{He}^+ + \mathrm{Li} \rightarrow \mathrm{He} + \mathrm{Li}^+$, which is the main dopant ionization channel in small He nanodroplets irradiated by XUV photons at any energy $h\nu > E_i$~\cite{Buchta:2013,LaForgePRL:2016,Ltaief:2020,asmussen2023dopant}.

\subsection{Total electron spectra}
\begin{figure}[t!]
	\center
    \includegraphics[width=1.0\columnwidth]{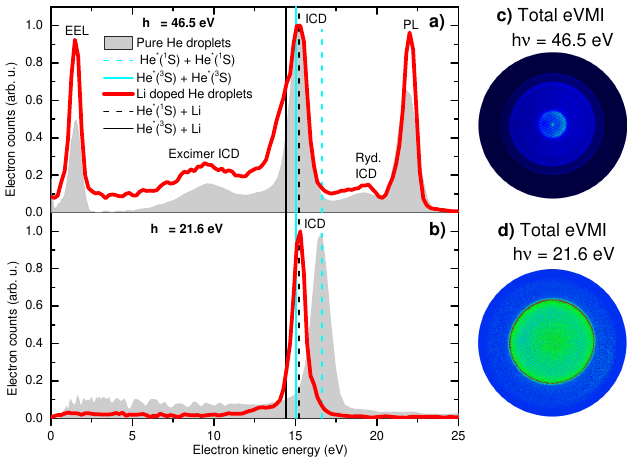}\caption{\label{fig:espec} a) Electron spectra measured for large He nanodroplets ($R=75$~nm) at $h\nu = 46.5$~eV [red line in a)] and $h\nu = 21.6$~eV [red line in b)]. The gray shaded areas in a) and b) represent reference spectra measured for pure large He nanodroplets at the same photon energies. Panels c) and d) depict velocity-map images of all emitted electrons from which the electron spectra plotted in red in a) and b) are inferred, respectively. The direction of incidence of the XUV photon beam is from right to left and the polarization points vertically in the paper plane. The dashed and solid vertical straight lines in cyan and black in a) and b) indicate characteristic electron energies for ICD a pair of He$^*$ atoms and a Li-He$^*$ atom pair. The photoline and electron energy-loss feature are denoted as PL and EEL, respectively.}
\end{figure}

Direct evidence for ICD of Li-doped large He nanodroplets producing both He and Li ions is obtained from electron spectra which are measured simultaneously with the mass spectra. The red line in Fig.~\ref{fig:espec}~a) shows the spectrum of all emitted electrons at $h\nu = 46.5$~eV. For comparison, Fig.~\ref{fig:espec}~b) shows the one measured in the regime of direct He nanodroplet photoexcitation at $h\nu = 21.6$~eV. Figs.~\ref{fig:espec}~c) and d) show velocity-map images of total electrons from which the electron spectra (red lines) in a) and b) are inferred, respectively. The velocity-map image in c) contains four distinguishable rings, whereas the one in d) consists of only one completely isotropic ring. Each ring converts into a peak in the electron spectra. The outermost and the innermost rings in c), which are due to directly emitted electrons and to photoelectrons that lost kinetic energy by inelastic scattering at surrounding He atoms, are slightly anisotropic with a preferred direction of emission to the right-hand side, \textit{i.~e.} towards the incident photon beam. This is due to a shadowing effect occurring in large He droplets which are opaque to both the XUV radiation and the emitted electrons~\cite{asmussen2023electron}. In contrast, the inner rings due to ICD are perfectly isotropic indicating complete redistribution of the He$^*$'s and He$_2^*$'s around the He droplet surface prior to ICD~\cite{ltaief:2023,ltaiefPRR:2024}.
The spectra recorded in the regime of resonant excitation at $h\nu = 21.6$~eV, shown in Fig.~\ref{fig:espec}~b), contain only one peak because ionization can occur only by the ICD reaction Eq.~(\ref{eq:He-ICD}) and, in case of Li-doped droplets, by Eq.~(\ref{eq:Li-ICD}). At $h\nu = 46.5$~eV, indeed the most prominent feature in the spectra of both pure and Li-doped He droplets is an ICD peak at the electron energy $E_e^\mathrm{ICD}\approx 15$~eV. Additional strong peaks are a photoline (PL) representing direct photoemission of He electrons with energy $E_e^\mathrm{PL} = h\nu - E_i = 21.9$~eV and an electron energy-loss (EEL) feature at $E_e^\mathrm{EEL}\approx 2$~eV resulting from inelastic scattering of photoelectrons at He atoms in the droplets~\cite{Shcherbinin:2019}. 
The gray shaded areas in Figs.~\ref{fig:espec}~a) and b) represent reference measurements with pure He nanodroplets of the same size. 

At $h\nu = 46.5$~eV, inelastic electron-He scattering mostly leads to He atoms in their lowest excited state, 1s2s\,$^3$S, with some contribution of 1s2s\,$^1$S and 1s2p\,$^3$P states~\cite{Shcherbinin:2019}. The latter is expected to rapidly relax into the 1s2s\,$^3$S state similarly to the 1s2p\,$^1$P droplet excitation that relaxes into the 1s2s\,$^1$S state within $\sim 0.5~$ps~\cite{Mudrich:2020,laforge2022relaxation}. Thus, we expect the EEL feature to peak at $E_e^\mathrm{EEL} = h\nu - E_i - E_\mathrm{3S} = 3.3$~eV and the He$^*$-ICD peak [Eq.~(\ref{eq:He-ICD})] to be centered at $E_e^\mathrm{ICD} = 2E_\mathrm{3S} - E_i = 15.0$~eV. Additionally, there are smaller features around $E_e = 9$ and $19$~eV which can be assigned to ionization of He$^*$ and possibly Li atoms by interaction with He$_2^*$ excimers and with He$^*$ in high-lying Rydberg states, respectively~\cite{ltaief:2023}. These are labeled `Excimer ICD' and `Ryd. ICD' in Fig.~\ref{fig:espec}~a), respectively. ICD involving He$^*$ in high Rydberg states is only observed in a narrow range of photon energies $h\nu = 44.5$-$47$~eV~\cite{ltaief:2023}. There, the photoelectron loses nearly all of its kinetic energy by inelastic scattering at a neighboring He atom and likely recombines with its parent He$^+$ ion before the latter stabilizes by forming a He$_2^+$ complex. The resulting energy of the ICD electron reflects the energy of the impact-excited He$^*$ atom interacting with the He$^*$ Rydberg atom, that is the energy of the 1s2s\,$^3$S state, 19.8~eV, reduced by the ionization energy of the Rydberg atom which is small. 

The excimer-ICD and Rydberg-ICD features are clearly absent in the electron spectrum measured at $h\nu = 21.6$~eV shown in Fig.~\ref{fig:espec}~b). The absence of Rydberg ICD at $h\nu = 21.6$~eV is evident as the excitation energy is insufficient to populate He$^*$ states higher than 1s2p\,$^1$P. The absence of excimer ICD in the spectra at $h\nu = 21.6$~eV indicates that He$_2^*$ excimers are formed, if at all, on a much longer timescale than ICD. As resonant photoexcitation leads to He$^*$ in the 1s2s\,$^1$S state, one may expect that He$_2^*$ excimers are formed in the $A\,^1\Sigma_u^+$ state by recombination of a He$^*$($^1$S) with a neighboring ground-state He atom. However, this process is hindered by a repulsive barrier in the He-He$^*$ pair potential~\cite{Buchenau:1991,Fiedler:2014}. In bulk liquid He, this reaction still occurs by many-body effects~\cite{nijjar2018conversion} on a time scale of $1.6~\mu$s, followed by spontaneous decay to the ground state within $<10$~ns~\cite{McKinsey:2003}. Indeed, these dynamics are much slower than ICD by reactions (\ref{eq:He-ICD}) and (\ref{eq:Li-ICD}) which is expected to take place within $\sim 1$~ps~\cite{LaForge:2021,Asmussen:2022} up to $\sim 1$~ns~\footnote{This estimate of the time constant is based on the assumption that two He$^*$'s emerging to the surface of He nanodroplets undergo a roaming motion before colliding and decaying by ICD. The roaming velocity is taken as the critical Landau velocity $\approx 60$~m/s~\cite{Brauer:2013} and the mean roaming distance is $\sim R$.}. Besides, vibrational relaxation of He$_2^*$ excimers bound to the surface of He droplets in high vibrational levels likely leads to their detachment from the droplets thereby quenching ICD involving He$_2^*$ excimers~\cite{Buchenau:1991}.

In contrast, photoelectron impact excitation of He$^*$ at $h\nu>45$~eV leads to a different mechanism of He$_2^*$ formation. The photoelectron is slowed down by the inelastic collision and localizes in a void bubble inside the He droplets. The photoion forms a He$_2^+$ which rapidly relaxes to its $v=0$ level within $\sim 100$~ps by strong coupling to the surrounding He~\cite{Benderskii:1999}. Due to the nearly identical shapes of the He$_2^+$ and He$_2^*$ potential energy curves at short distance, electron-He$_2^+$ recombination efficiently populates He$_2^*$ in the lowest vibrational level $v=0$ of the long-lived $a\,^3\Sigma_u^+$ state. In Li-doped large He droplets, He$_2^*$ excimers can decay by ICD involving either another He$_2^*$ excimer, a He$^*$ atom, or a Li atom or Li cluster by the reactions
\begin{eqnarray}
\label{eq:He2-ICD}
\mathrm{He}_2^* + \mathrm{He}_n^* &\rightarrow& \mathrm{He}_2 + \mathrm{He}_n^+ + e_\mathrm{ICD},\\
\label{eq:He2Li-ICD}
\mathrm{He}_2^* + \mathrm{Li}_m &\rightarrow& \mathrm{He}_2 + \mathrm{Li}_m^+ + e_\mathrm{ICD}.
\end{eqnarray}
Here, $n=1,\,2$ and $m$ is the number of Li atoms per Li cluster.

\begin{figure*}
	\centering
    \includegraphics[width=0.95\textwidth]{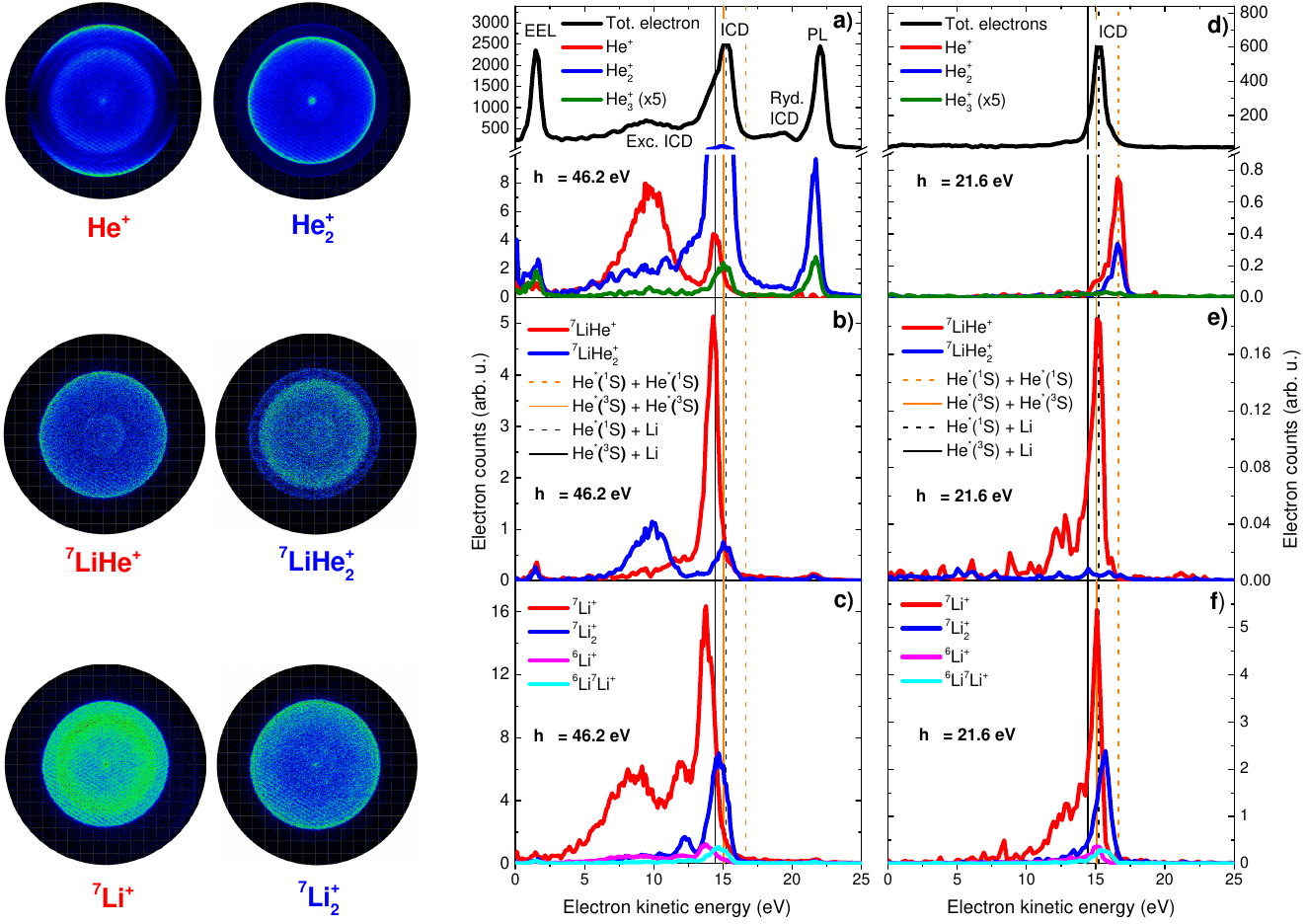}\caption{\label{fig:coincSpec} Spectra of all electrons (black lines) and of electrons recorded in coincidence with specific ions (colored lines) at $h\nu = 46.2$~eV (left column) and at $h\nu = 21.6$~eV (right column) for large He nanodroplets ($R= 75$~nm) doped with Li atoms ($T_\mathrm{Li}= 400~$$^\circ$C). Typical electron VMIs that are used to infer the electron spectra measured in coincidence with He$^+$ [red line in a)], He$_2^+$ [blue line in a)], LiHe$^+$ [red line in b)] and LiHe$_2^+$ [blue line in b)], $^7$Li$^+$ [red line in c)] and $^7$Li$_2^+$ [blue line in c)] are shown from the top to the bottom in the left hand side of panels a-c), respectively. All spectra are normalized to the photon flux and to the $h\nu$-dependent absorption cross section of He. Vertical straight lines indicate ICD electron energies expected based on atomic term values and ionization energies.}
\end{figure*}
\subsection{Electron spectra measured in coincidence with He and Li ions}
In the spectrum of Li-doped He droplets shown in Fig.~\ref{fig:espec}~a), the main ICD peak at $E_e = 15$~eV has a shoulder towards lower electron energies which is not present in the reference spectrum of pure droplets (gray shaded area). This is due to Li ICD [Eq.~(\ref{eq:Li-ICD})] producing electrons with a characteristic energy $E_e^\mathrm{Li\mhyphen ICD} = E_\mathrm{3S} - E_i^\mathrm{Li} = 14.4$~eV, see the solid black vertical line in Fig.~\ref{fig:espec}. A direct Li-ICD signal is obtained when measuring photoelectron spectra in coincidence with Li ions, see Fig.~\ref{fig:coincSpec} and SM Fig. 1. The left column of panels in Fig.~\ref{fig:coincSpec} show spectra recorded at $h\nu = 46.2$~eV. Spectra recorded in coincidence with He ions are shown in panel a), those recorded in coincidence with [LiHe]$^+$ and [LiHe$_2$]$^+$ ions are shown in b) and the Li$^+$-correlated electron spectra are shown in c). 

Typical velocity map electron images used to infer the electron spectra measured in coincidence with He$^+$, He$_2^+$, [LiHe]$^+$, [LiHe$_2$]$^+$, Li$^+$ and Li$_2^+$ are shown on the left hand side of Fig.~\ref{fig:coincSpec}~a-c), respectively. The outermost ring in the image measured in coincidence with He$^+$, which is due direct photoemission, is anisotropic with a preferred direction of emission along the polarization axis of the XUV light. This is expected as He$^+$ ions produced by photoionization stem from free He atoms accompanying the droplet beam, whereas photoionization of He droplets produces He$_2^+$ and larger cluster ions~\cite{Shcherbinin:2019}. The outer photoelectron ring in the He$_2^+$ coincidence image and the innermost EEL ring in all images except the one for He$^+$ show a slight forward-backward asymmetry due to a shadowing effect as discussed above~\cite{asmussen2023electron}. In contrast, all inner rings representing ICD electrons are perfectly isotropic indicating complete loss of directional information imprinted by the photoabsorption process. This indicates the complete redistribution of the He$^*$ and He$_2^*$ around the droplet surface prior to ICD, implying that this type of ICD is a slow process~\cite{ltaief:2023,ltaiefPRR:2024}. 

The panels on the right-hand side show the same coincidences recorded at $h\nu = 21.6$~eV. These spectra have been discussed in detail earlier~\cite{Ltaief:2019,ltaiefPRR:2024}; they essentially feature one narrow peak due to the either ICD of Li [Eq.~(\ref{eq:Li-ICD})] or He$^*$ in the 1s2s\,$^1$S state [Eq.~(\ref{eq:He-ICD})]. The fact that the ICD peak in the spectrum of all electrons [top of panel d) and gray shaded area in Fig.~\ref{fig:espec}~b)] matches the electron energy ($E_e=15.2$~eV) corresponding to ICD of Li interacting with He$^*$(1s2s\,$^1$S), see the dashed black vertical line, indicates that Li ICD is significantly more abundant than ICD involving two He$^*$'s [Eq.~(\ref{eq:He-ICD})], see also the mass spectrum (Fig.~\ref{fig:massspec}) where the Li$^+$ peak is the highest. 

All spectra recorded at $h\nu = 46.2$~eV [Fig.~\ref{fig:coincSpec} a)-c)] contain the same features present in the spectrum of all electrons described above [Fig.~\ref{fig:espec}~a)] but with variable amplitudes. The spectra taken in coincidence with He$_n^+$, $n=1,\,2,\,3$ have been discussed before~\cite{ltaief:2023}. Note the surprising, counter-intuitive finding that ICD involving the decay of He$_2^*$ excimers primarily produce He$^+$ atomic ions whereas He$_2^+$ dimer ions are primarily produced by ICD of He$^*$ atoms [Eq.~(\ref{eq:He-ICD})]. We interpreted the preferred ejection of He$^+$ upon ICD involving He$_2^*$ by the dissociation of He$_2^*$ into two energetic neutral He atoms where one of them subsequently collides with the nascent He$^+$ ICD ion. In this way, the He$^+$ is ``kicked out of the droplet'' whereas He$^+$ formed with low kinetic energy by He$^*$ ICD tends to bind a neighboring He atom before leaving the droplet as a He$_2^+$ molecular ion similar to the formation of He$_2^+$ ions by direct photoionization, see~\cite{BuchtaJCP:2013} and the supplementary material of Ref.~\cite{ltaief:2023}.

A similar trend is observed for the Li$^+$ and Li$_2^+$ ions, see Fig.~\ref{fig:coincSpec}~c). All the electron spectra measured in coincidence with Li$^+$ and Li$_2^+$ ions exhibit three main ICD features. The ICD peaks in the atomic Li$^+$ coincidence spectra appear around $E_e =8.6$, 12.0 and 13.8~eV. The ICD peaks in the Li$_2^+$ coincidence spectra appear slightly shifted up in energy to $E_e =8.9$, 12.3 and 14.7~eV. This up-shift is probably due to the lower ionization energy of Li clusters as compared to their constituent atoms. The Li-ICD features peaking around $E_e =13.8$ and 14.7~eV result from the ICD reaction (2) which involves the decay of He*($^3$S) metastable atoms interacting with Li atoms and Li dimers, respectively. The broad ICD features peaking at $E_e =8.6$ and 8.9~eV in the electron spectra measured in coincidence with Li$^+$ and Li$_2^+$ are due to ICD involving the decay of He$_2^*$ excimers interacting with Li atoms and Li dimers, respectively. The additional small maxima in the Li$^+$ and Li$_2^+$ coincidence electron spectra in c) around $E_e = 12.0$ and $12.3$~eV are yet unassigned. 

Note that the main ICD peak at $E_e^\mathrm{Li\mhyphen ICD} = 13.8$~eV in the electron spectrum of Li$^+$ is down-shifted by 0.6~eV with respect to the value expected for Li+He$^*$($^3$S) ICD assuming atomic He$^*$ excitation and Li ionization energies. Such a down-shift has previously been measured at $h\nu = 21.6$~eV for Li-doped He nanodroplets and was interpreted by the binding energy of the LiHe$^*$ complex prior to ICD occuring at short interatomic distance~\cite{Ltaief:2019}. As the binding energy is converted into kinetic energy of the He and Li atoms which is conserved during the ICD reaction, the Li$^+$ ICD ion is formed with substantial kinetic energy which facilitates its ejection from the droplet. The $^1$S-ICD peak measured in coincidence with Li$^+$ at $h\nu = 21.6$~eV, shown in Fig.~\ref{fig:coincSpec}~f), appears barely down-shifted, contrary to previous measurements~\cite{Ltaief:2019}, presumably due to the larger size of the He droplets used in this work that might cause a lowering of the Li ionization energy.


Only the electron spectra measured in coincidence with Li$_2^+$ [blue and cyan curves in Fig.~\ref{fig:coincSpec} f)] show a clear shift of the main ICD peak to higher kinetic energy due to lowering of the ionization energy of Li$_2$ or Li clusters with respect to Li atoms. On the other hand, ICD occurring between Li and He$^*$($^1$S) or He$^*$($^3$S) atoms at large distances should lead to a nearly unshifted ICD peak, as it has been observed before for small He nanodroplets by detecting all emitted electrons~\cite{Ltaief:2019}. This can also be seen in the all-electron spectra shown in Fig.~\ref{fig:coincSpec}~d) and Fig.~\ref{fig:espec}~b). Here, the Li$^+$ is formed with vanishing kinetic energy leading to its sinking into the droplet, thus evading its detection~\cite{Ltaief:2019}. 

Li$^+$ ions produced by the ICD reaction (2) can bind one He atom before being ejected from the He nanodroplet thereby forming a [LiHe]$^+$ complex. Typical electron spectra measured in coincidence with [LiHe]$^+$ ions at $h\nu = 46.2$ and $21.6$~eV are shown in Fig.~\ref{fig:coincSpec}~b) and e), respectively. Both spectra exhibit only one main ICD peak which is nearly unshifted with respect to the atomic energy values. 
For the [LiHe]$^+$-electron coincidence spectrum measured at $h\nu = 46.2$~eV, one would expect to observe a feature due to He$_2^*$-induced ICD peaking around $E_e = 9$~eV as well. The absence of this feature is likely related to the dissociation of the He$_2^*$ excimer to energetic ground-state He atoms upon ICD, as discussed above. Transfer of kinetic energy from one of the energetic He atoms to the nascent Li$^+$ ion hampers the formation of [LiHe]$^+$. This `kicked' Li$^+$ ion leaves the droplet as free Li$^+$ whereas Li$^+$ ions produced by reaction (2) tend to bind a He atom before being ejected.

In contrast, [LiHe$_2$]$^+$ ions are formed by a different mechanism. Before their decay, some of the He$_2^*$'s created by electron-He$_2^+$ ion recombination can instead bind Li atoms when they reach the surface of the He nanodroplet, thereby leading to the formation of LiHe$_2^*$ complexes. The latter can in turn get ionized by ICD through interaction with He$_2^*$ or He$^*$ leading to [LiHe$_2$]$^+$ ions and an ICD electron with kinetic energy around 10 or 15.2~eV, respectively [see the blue curve in Fig.~\ref{fig:coincSpec}~b)]. Interestingly, no such features are observed in the electron spectrum measured at $h\nu = 21.6$~eV [blue line in Fig.~\ref{fig:coincSpec} e)]. The absence of [LiHe$_2$]$^+$ ions at $h\nu = 21.6$~eV is a further indication that the formation of He$_2^*$ out of He$^*$($^1$S) and He atoms does not occur on the timescale of ICD. 

\begin{figure}[t!]
	\center
    \includegraphics[width=1.0\columnwidth]{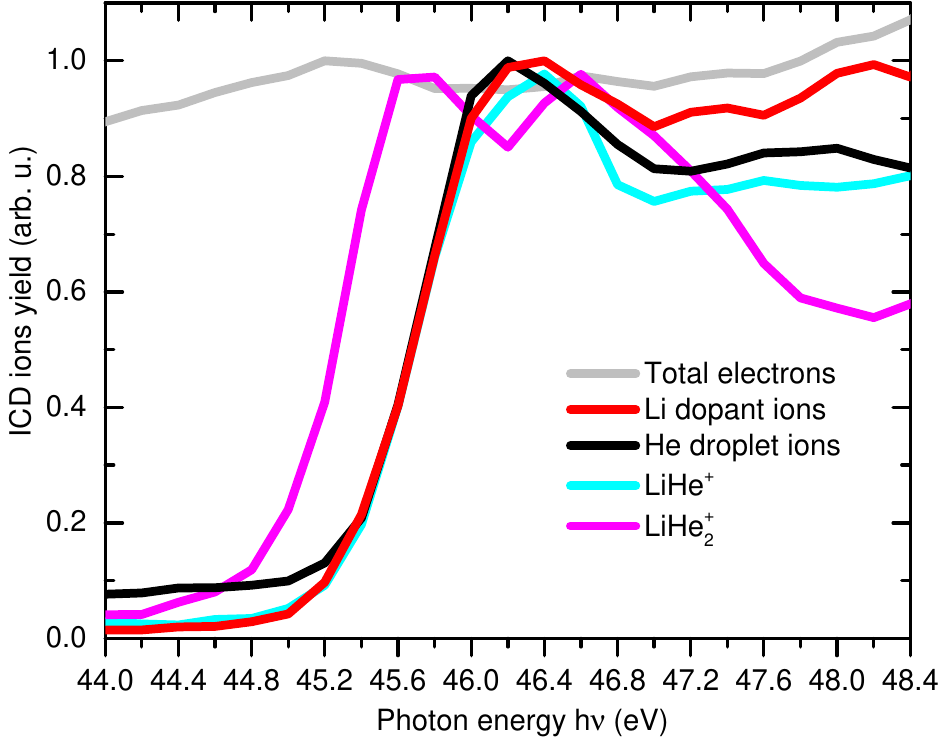}\caption{\label{fig:hv-scanions}
    Yield of all ions containing Li (red line) and sum of He$_n^+$ ion yields for $n=1$-3 (black line) as a function of photon energy. Additionally, the [LiHe]$^+$ and [LiHe$_2$]$^+$ yields are shown as cyan and magenta lines.
    }
\end{figure}
\subsection{Photon-energy dependence}
To get an overview of the ICD efficiency across the photoionization spectrum of He droplets near the threshold for electron-impact excitation, the photon energy was tuned in the range $h\nu = 44$-$48.4$~eV while measuring the yield of Li ions, He ions, and all electrons, see Fig.~\ref{fig:hv-scanions}. The yield of characteristic ICD electrons is shown in SM Fig.~2-4). All the mass-specific electron and ion yields show a clear onset of ICD around $h\nu = 45$~eV and a maximum around 46~eV. However, in contrast to the ICD electron yields, all the ion yields exhibit a constant signal level below their onset, especially the He ions (black line in Fig.~\ref{fig:hv-scanions}). This is due to direct photoionization of the He nanodroplets producing He ions and by electron-transfer ionization producing Li ions. All of the ion and electron yields follow essentially the same behavior except that of all electrons and of [LiHe$_2$]$^+$ ions. Surprisingly, the yield of [LiHe$_2$]$^+$ ions exhibits an earlier onset at $h\nu = 44.4$~eV and a maximum at $h\nu = 45.2$~eV, whereas the onset and the maximum of all other yields is at $h\nu = 45.2$~eV and $h\nu = 46.2$~eV, respectively. This energy difference of about 0.8~eV matches exactly the energy shift of the He electron impact excitation energies in He nanodroplets with respect to atomic-level energies, as observed previously for pure large He nanodroplets~\cite{ltaief:2023}. We tentatively interpret the missing shift of the onset of [LiHe$_2$]$^+$ ions and electrons by the site-selective formation of LiHe$_2^*$ at the droplets surface where the He$^*$ excitation and the decay is unperturbed by the He environment. 

In contrast to all mass-specific electron and ion yields, the yield of all electrons remains almost constant in the photon energy range $h\nu = 44$-$48.4$~eV. 
This implies that the opening of the new ICD channel by photoelectron impact excitation of He$^*$ and subsequent ICD of He$^*$ or Li does not enhance the total yield of electrons and ions; it merely leads to a redistribution of electrons from the photoline to the lower-energy electron-energy-loss (EEL) and ICD features. This can be nicely seen in SM Fig. 5. The fact that every photoelectron that undergoes impact excitation and thus is missing in the photoline reappears as an EEL or ICD electron implies that ICD proceeds with nearly unity probability.

\begin{figure}[t!]
	\center
    \includegraphics[width=0.85\columnwidth]{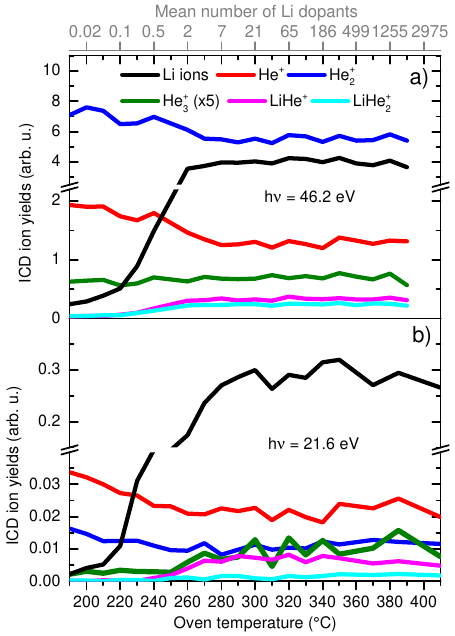}\caption{\label{fig:OvenT} Yields of He and Li ions as a function of the temperature of the Li doping cell measured for Li-doped large He nanodroplets ($R=75$~nm) at $h\nu=46.2~$eV a) and at $h\nu=21.6~$eV b). The yield of Li ions plotted in black is the sum of $^7$Li$^+$, $^7$Li$_2^+$, $^6$Li$^+$ and $^6$Li$_2^+$ yields. All the yield curves are normalized to the photon flux and to the $h\nu$-dependent absorption cross section of He. The estimated mean number of Li atoms picked up by the droplets is shown at the top of panel a).}
\end{figure}
\subsection{Dependence on the doping level}
To further characterize this new indirect ICD process of dopants attached to large He nanodroplets, we varied the temperature $T_\mathrm{Li}$ of the Li doping cell and repeatedly recorded mass spectra. Fig.~\ref{fig:OvenT} shows the yields of Li$^+$ and He$^+$ ions for $h\nu=46.2~$eV in panel a). For reference, panel b) shows the same ion yields recorded at $h\nu=21.6~$eV, where He droplets are directly photoexcited at the 1s2p resonance. 
At both photon energies we observe a concurrent increase of the yield of all Li-containing ions and a drop of the yield of He$^+$ and He$_2^+$ ions in the range from $T_\mathrm{Li}=190$ to $270\,^\circ$C. In this temperature range, the probability of He nanodroplets to be doped with one or more Li atoms increases from nearly zero to about unity. At higher cell temperatures up to 430$\,^\circ$C, the ion yields remain nearly constant. In this range, nearly all He droplets are doped with more than one Li atom which form clusters at the droplet surface~\cite{Buenermann:2011}. Most likely clustering occurs into multiple centers as previously observed for other metals~\cite{ernst2021metal}, but from our data we cannot conclude on this point. The weak dependence of the ion yields on the doping level at higher temperatures $T_\mathrm{Li}>270\,^\circ$C indicates that the probability of dopant ionization is independent of the size and number of the dopant clusters formed on the droplets. The slight drop of the He ion yields in the range of rising Li ion yields indicates that ionization and ejection of Li and He ions are competing processes; both are created either directly through photoionization (followed by charge transfer ionization in the case of Li) or, to a larger extent, indirectly by ICD by reactions~(\ref{eq:Li-ICD}) and (\ref{eq:He-ICD}). Clearly, depletion of the He droplet beam due to full evaporation and scattering of the droplets out of the beam~\cite{Buenermann:2011} does not play any role here given the large size of the droplets.

\begin{figure}[t!]
	\center
    \includegraphics[width=0.95\columnwidth]{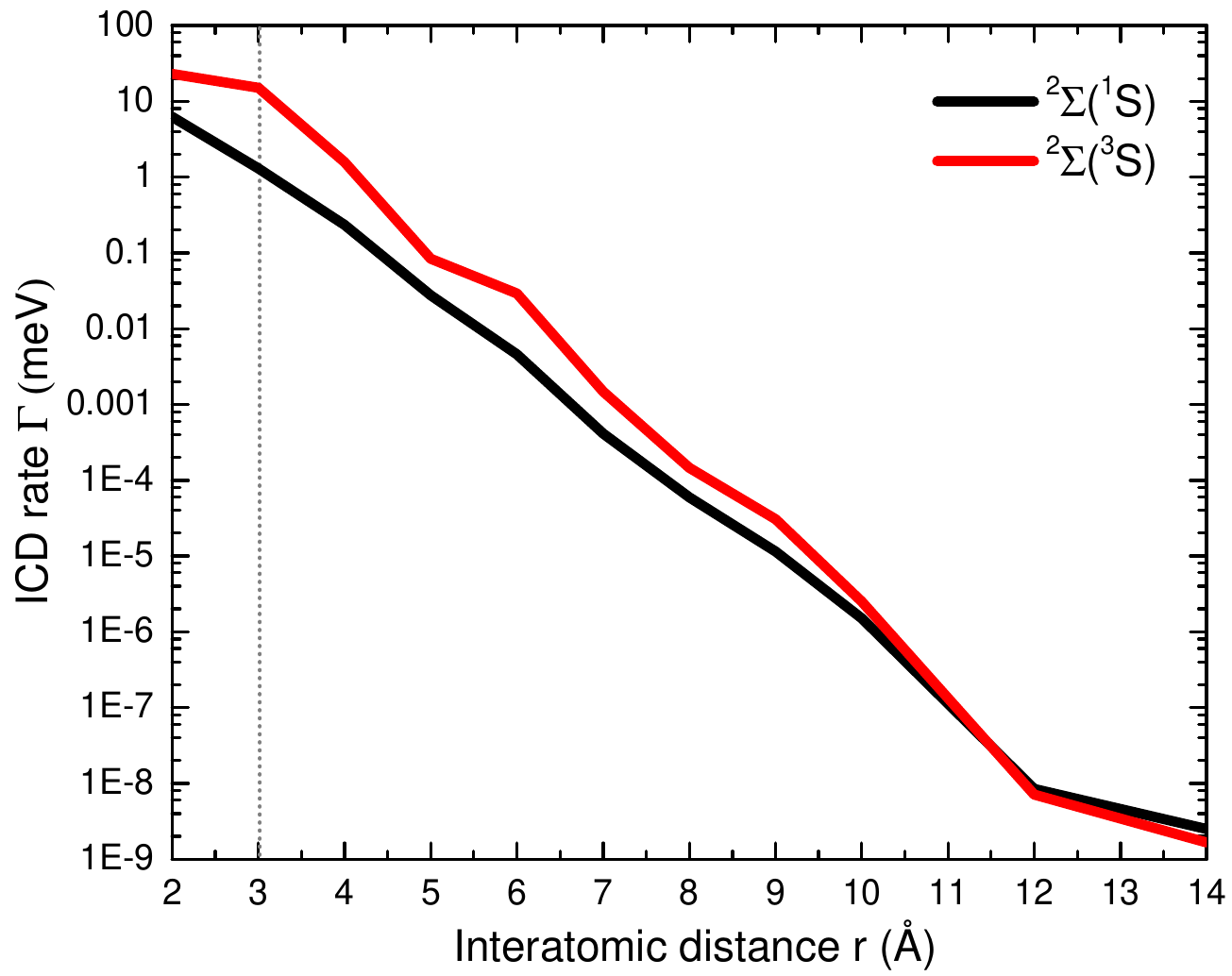}\caption{\label{fig:ICDrate} Calculated average ICD rates for the He$^*$+Li reaction involving the two lowest excited singlet and triplet states of He$^*$, 1s2s\,$^{1,3}\Sigma$. The vertical dashed line indicates the interatomic distance corresponding to the minimum of the potential energy curve of He$^*$Li~\cite{movre:2000}.}
\end{figure}
The main difference between the yields measured at the two photon energies, aside from the overall much higher values at $h\nu=46.2~$eV, is the higher yield of He$_2^+$ compared to Li ions at $h\nu=46.2~$eV in the full range of Li doping ($T_\mathrm{Li}>270\,^\circ$C), whereas at $h\nu=21.6~$eV the Li ion yield is by far the highest. Upon resonant excitation of the He droplets at $h\nu=21.6~$eV, only Li ions can be produced in singly-excited He nanodroplets; the formation of He ions by ICD [Eq.~(\ref{eq:He-ICD})] requires two or more photons to be absorbed by one droplet which is less likely and only occurs in large droplets. In contrast, at $h\nu=46.2~$eV both the formation of He and Li ions by ICD requires large droplets where electron impact excitation of He$^*$ and electron-ion recombination is efficient. Another reason for the high efficiency of the ICD reactions (\ref{eq:Li-ICD}) and (\ref{eq:He-ICD}) is that photoexcitation and electron-impact excitation produce He$^*$'s in different states; following resonant photoexcitation, He nanodroplets rapidly relax into the localized 1s2s\,$^1$S metastable singlet state~\cite{Mudrich:2020}. In contrast, electron impact excitation populates He$^*$'s in both singlet and triplet states, whereby the latter relax into the lowest 1s2s\,$^3$S localized state. To some extent also He$_2^*$ excimers are formed, likely in their lowest $^3\Sigma$ ($v=0$) state. Both He$^*$ and He$_2^*$ in triplet states have very long lifetimes (15~$\mu$s and 13~s, respectively~\cite{McKinsey:1999,McKinsey:2003,Keto:1974,Buchenau:1991}) which allows these excitations to accumulate in one droplet over multiple pulses of the synchrotron radiation that come at a repetition period of $9.5$~ns. In contrast, He$^*$ and He$_2^*$ excited in the singlet state have shorter lifetimes $<10$~ns and 1.6~$\mu$s, respectively~\cite{McKinsey:2003,carter:2017} and therefore partly decay by fluorescence emission before ICD occurs. This can also explain the absence of the signature of ICD out of $^1$S singlet states in the electron spectra presented in Figs.~\ref{fig:espec} a) and \ref{fig:coincSpec}~a-c).

Note that the contribution of $^1$S states in the ICD electron signal is very small and was observed only at higher photon energies where the cross section for electron impact excitation leading to He$^*$'s in singlet states exceeds that for triplet excitation, \textit{i.~e.} at $h\nu\gtrsim 45.0~$eV (see SM Fig.~6). According to the electron-impact excitation cross sections of He atoms~\cite{vcermak1966individual,dugan1967excitation,Ralchenko:2008}, above the electron impact excitation threshold of the 1s2s\,$^1$S state, triplet states are expected to be formed about twice as efficiently as singlet states.

Furthermore, the ICD rates of Li and He$^*$ atoms in singlet and triplet states may also differ. To quantitatively investigate this, we have carried out \textit{ab initio} calculations of the ICD rates for He$^*$ in the $^1$S and the $^3$S states interacting with a Li atom at various interatomic distances $r$ using the Fano–CI–Stieltjes method~\cite{miteva:2017,Ltaief:2019}. The resulting ICD rates, shown in Fig.~\ref{fig:ICDrate}, exhibit an exponential decay as a function of $r$, indicative of the charge-exchange mechanism~\cite{Ltaief:2019}. The ICD rate of Li and He$^*$ in the $^3$S state is increasingly higher than that of the $^1$S state for decreasing $r<10$~\AA; at the minimum of the He$^*$-Li binding potential, $r=3$~\AA, where ICD is likely to occur, the $^3$S ICD rate exceeds that of the $^1$S state by about a factor of 5. We take this factor as an additional reason for the low contribution of $^1$S ICD in the electron spectra measured in the regime of electron-impact excitation.

\begin{figure}[t!]
	\center
    \includegraphics[width=0.85\columnwidth]{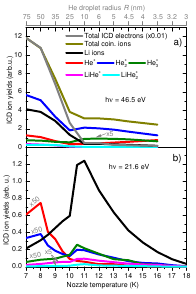}\caption{\label{fig:NozzleT}
    Yields of He and Li ions as a function of the temperature of the He nozzle measured at $h\nu=46.5~$eV a) and at $h\nu=21.6~$eV b). The Li oven temperature $T_\mathrm{Li}$ was set to 400\,$^\circ$C. All the curves are normalized to the photon flux and to the $h\nu$-dependent absorption cross section of He. The gray axis shown in the top of panel a) indicates the He droplet radius $R$ determined according to~Ref.~\cite{Toennies:2004}.}
\end{figure}
\subsection{Dependence on the He nanodroplet size}
The dependence of the measured ion yields on the nozzle temperature which determines the He droplet size is shown in Fig.~\ref{fig:NozzleT}. Here, we see significant differences between the yields measured at $h\nu=46.5~$eV and at $h\nu=21.6~$eV shown in Fig.~\ref{fig:NozzleT}~a) and~b), respectively. In particular, at $h\nu=21.6~$eV the yield of Li ions, which is by far the highest of all ions, features a pronounced maximum around the critical point at $T_\mathrm{He}=11$~K where the mean He droplet size sharply increases from $N_\mathrm{He}\approx 2\times 10^4$ He atoms per droplet ($R= 6$~nm) to $N_\mathrm{He}\approx 10^5$ ($R= 10$~nm)~\cite{Toennies:2004}. These conditions seem to be optimal in terms of high levels of Li doping given the rather large droplet size and a high number density of droplets in the interaction volume. In contrast, at $h\nu=46.5~$eV the Li ion yield is low at higher nozzle temperatures and only increases as the nozzle is cooled to $T_\mathrm{He}<11~$K. The electron-impact excitation and electron-ion recombination mechanisms are active only for He droplets larger than a critical size $R\approx 10$~nm~\cite{ltaief:2023}. Note that at $h\nu=46.5~$eV the yield of He$_2^+$ is significantly higher than that of He$^+$, whereas at $h\nu=21.6~$eV the He$^+$ yield exceeds that of He$_2^+$ at $T_\mathrm{He}<10~$K. This is likely due to the different shapes of the potential energy curves of He$^*$-He$^*$ for He$^*$'s in singlet and in triplet states~\cite{ltaief:2023,LaForge:2021}. The interaction potential of two He$^*$($^1$S) atoms excited at $h\nu=21.6~$eV features an extended minimum at relatively large interatomic distance (4~\AA) at which ICD leads to a nearly unbound He-He$^+$ pair. The decay of a pair of impact-excited He$^*$($^3$S) atoms at $h\nu=46.5~$eV occurring at the potential minimum at 3.5~\AA~leads to more deeply bound He$_2^+$.

In addition to the yields of individual ions, Fig.~\ref{fig:NozzleT}~a) shows the sum of all ions (measured in coincidence with an electron) as a yellow line and the yield of all ICD electrons as a gray line. In the regime of small He droplets, $T_\mathrm{He}>11~$K, ions and electrons are produced to some extent by direct photoionization. Surprisingly, no Li-ICD electrons are observed in this regime although an electron-energy loss feature is visible in the electron spectra indicating the excitation of He$^*$ atoms in the droplets~\cite{Shcherbinin:2019}. It appears that He$^*$'s created by electron-impact excitation in small He nanodroplets are mostly ejected from the droplets instead of decaying by ICD. As the nozzle is cooled to $T_\mathrm{He}<11~$K where the droplets grow much larger, a sharp increase of both ion and electron yields is measured and the emission of electrons and ions is dominated by ICD. 

\section{Conclusion}
We have unraveled the intricate ionization dynamics of large He nanodroplets doped with lithium atoms which is initiated by XUV irradiation far above the ionization threshold of He. Various ICD processes involving metastable He$^*$ atoms, He$^*$ in high-lying Rydberg states, and He$_2^*$ excimers are identified, all of which are induced by inelastic scattering of the photoelectron and electron-ion recombination in the droplets. These indirect ICD processes are found to be even more efficient in producing dopant electrons and ions than ICD following direct resonant photoexcitation of the droplets. In the inelastic-scattering regime, He$^*$ and He$_2^*$ are formed in metastable triplet states which decay by ICD to produce He and Li ions and characteristic electrons. In contrast, in the regime of resonant photoexcitation only ICD of He$^*$'s in the lowest excited singlet state was observed. This finding confirms the crucial role of recombination of the inelastically scattered photoelectron with its parent ion, which we have conjectured earlier~\cite{ltaief:2023,ltaiefPRR:2024}. The associated sharp, nearly unperturbed features in the electron spectra and the efficient ICD of lithium residing at the droplet surface indicate that the He$^*$ and He$_2^*$ fully relax internally and externally by emerging to the droplet surface prior to ICD. We speculate that the time scale of this type of ICD is determined by the dynamics of electron-ion recombination ($\sim 100$~ps~\cite{Benderskii:1999}) and of the roaming of the He$^*$ and He$_2^*$ excitations about the droplet surface ($\sim 0.1$-1~ns). Future pump-probe experiments similar to those performed in the regime of resonant droplet excitation~\cite{LaForge:2021,Asmussen:2022} will give further insight into these intriguing indirect autoionization processes.

\section{Acknowledgement}
M.M. and L.B.L. acknowledge financial support by the Danish Council for Independent Research Fund (DFF) via Grant No. 1026-00299B. L. B. L acknowledges support by the Villum foundation via the Villum Experiment grant No. 58859. SRK thanks Dept. of Science and Technology, Govt. of India, for support through the DST-DAAD scheme and Science and Eng. Research Board. SRK and KS acknowledge the support of the Scheme for Promotion of Academic Research Collaboration, Min. of Edu., Govt. of India, and the Institute of Excellence programme at IIT-Madras via the Quantum Center for Diamond and Emergent Materials. SRK gratefully acknowledges support of the Max Planck Society's Partner group programme. NS and SRK gratefully acknowledge CEFIPRA (Indo-French Centre for the Promotion of Advanced Research) for generous support. The research leading to this result has been supported by the COST Action CA21101 ``Confined Molecular Systems: From a New Generation of Materials to the Stars (COSY)''.

\bibliography{ROPP_Bib.bib}

\begin{thebibliography}{70}%
\makeatletter
\providecommand \@ifxundefined [1]{%
 \@ifx{#1\undefined}
}%
\providecommand \@ifnum [1]{%
 \ifnum #1\expandafter \@firstoftwo
 \else \expandafter \@secondoftwo
 \fi
}%
\providecommand \@ifx [1]{%
 \ifx #1\expandafter \@firstoftwo
 \else \expandafter \@secondoftwo
 \fi
}%
\providecommand \natexlab [1]{#1}%
\providecommand \enquote  [1]{``#1''}%
\providecommand \bibnamefont  [1]{#1}%
\providecommand \bibfnamefont [1]{#1}%
\providecommand \citenamefont [1]{#1}%
\providecommand \href@noop [0]{\@secondoftwo}%
\providecommand \href [0]{\begingroup \@sanitize@url \@href}%
\providecommand \@href[1]{\@@startlink{#1}\@@href}%
\providecommand \@@href[1]{\endgroup#1\@@endlink}%
\providecommand \@sanitize@url [0]{\catcode `\\12\catcode `\$12\catcode `\&12\catcode `\#12\catcode `\^12\catcode `\_12\catcode `\%12\relax}%
\providecommand \@@startlink[1]{}%
\providecommand \@@endlink[0]{}%
\providecommand \url  [0]{\begingroup\@sanitize@url \@url }%
\providecommand \@url [1]{\endgroup\@href {#1}{\urlprefix }}%
\providecommand \urlprefix  [0]{URL }%
\providecommand \Eprint [0]{\href }%
\providecommand \doibase [0]{https://doi.org/}%
\providecommand \selectlanguage [0]{\@gobble}%
\providecommand \bibinfo  [0]{\@secondoftwo}%
\providecommand \bibfield  [0]{\@secondoftwo}%
\providecommand \translation [1]{[#1]}%
\providecommand \BibitemOpen [0]{}%
\providecommand \bibitemStop [0]{}%
\providecommand \bibitemNoStop [0]{.\EOS\space}%
\providecommand \EOS [0]{\spacefactor3000\relax}%
\providecommand \BibitemShut  [1]{\csname bibitem#1\endcsname}%
\let\auto@bib@innerbib\@empty
\bibitem [{\citenamefont {Jahnke}\ \emph {et~al.}(2020)\citenamefont {Jahnke}, \citenamefont {Hergenhahn}, \citenamefont {Winter}, \citenamefont {Do{\"o}rner}, \citenamefont {Fru{\"u}hling}, \citenamefont {Demekhin}, \citenamefont {Gokhberg}, \citenamefont {Cederbaum}, \citenamefont {Ehresmann}, \citenamefont {Knie} \emph {et~al.}}]{Jahnke:2020}%
  \BibitemOpen
  \bibfield  {author} {\bibinfo {author} {\bibfnamefont {T.}~\bibnamefont {Jahnke}}, \bibinfo {author} {\bibfnamefont {U.}~\bibnamefont {Hergenhahn}}, \bibinfo {author} {\bibfnamefont {B.}~\bibnamefont {Winter}}, \bibinfo {author} {\bibfnamefont {R.}~\bibnamefont {Do{\"o}rner}}, \bibinfo {author} {\bibfnamefont {U.}~\bibnamefont {Fru{\"u}hling}}, \bibinfo {author} {\bibfnamefont {P.~V.}\ \bibnamefont {Demekhin}}, \bibinfo {author} {\bibfnamefont {K.}~\bibnamefont {Gokhberg}}, \bibinfo {author} {\bibfnamefont {L.~S.}\ \bibnamefont {Cederbaum}}, \bibinfo {author} {\bibfnamefont {A.}~\bibnamefont {Ehresmann}}, \bibinfo {author} {\bibfnamefont {A.}~\bibnamefont {Knie}}, \emph {et~al.},\ }\bibfield  {title} {\bibinfo {title} {Interatomic and intermolecular coulombic decay},\ }\href@noop {} {\bibfield  {journal} {\bibinfo  {journal} {Chem. Rev.}\ }\textbf {\bibinfo {volume} {120}},\ \bibinfo {pages} {11295} (\bibinfo {year} {2020})}\BibitemShut {NoStop}%
\bibitem [{\citenamefont {Cederbaum}\ \emph {et~al.}(1997)\citenamefont {Cederbaum}, \citenamefont {Zobeley},\ and\ \citenamefont {Tarantelli}}]{Cederbaum:1997}%
  \BibitemOpen
  \bibfield  {author} {\bibinfo {author} {\bibfnamefont {L.~S.}\ \bibnamefont {Cederbaum}}, \bibinfo {author} {\bibfnamefont {J.}~\bibnamefont {Zobeley}},\ and\ \bibinfo {author} {\bibfnamefont {F.}~\bibnamefont {Tarantelli}},\ }\bibfield  {title} {\bibinfo {title} {Giant intermolecular decay and fragmentation of clusters},\ }\href@noop {} {\bibfield  {journal} {\bibinfo  {journal} {Phys. Rev. Lett.}\ }\textbf {\bibinfo {volume} {79}},\ \bibinfo {pages} {4778} (\bibinfo {year} {1997})}\BibitemShut {NoStop}%
\bibitem [{\citenamefont {Hergenhahn}(2011)}]{Hergenhahn:2011}%
  \BibitemOpen
  \bibfield  {author} {\bibinfo {author} {\bibfnamefont {U.}~\bibnamefont {Hergenhahn}},\ }\bibfield  {title} {\bibinfo {title} {Interatomic and intermolecular coulombic decay: The early years},\ }\href@noop {} {\bibfield  {journal} {\bibinfo  {journal} {J. Electron. Spectrosc. Relat. Phenom.}\ }\textbf {\bibinfo {volume} {184}},\ \bibinfo {pages} {78 } (\bibinfo {year} {2011})}\BibitemShut {NoStop}%
\bibitem [{\citenamefont {Stumpf}\ \emph {et~al.}(2016)\citenamefont {Stumpf}, \citenamefont {Gokhberg},\ and\ \citenamefont {Cederbaum}}]{Stumpf:2016}%
  \BibitemOpen
  \bibfield  {author} {\bibinfo {author} {\bibfnamefont {V.}~\bibnamefont {Stumpf}}, \bibinfo {author} {\bibfnamefont {K.}~\bibnamefont {Gokhberg}},\ and\ \bibinfo {author} {\bibfnamefont {L.~S.}\ \bibnamefont {Cederbaum}},\ }\bibfield  {title} {\bibinfo {title} {The role of metal ions in x-ray-induced photochemistry},\ }\href@noop {} {\bibfield  {journal} {\bibinfo  {journal} {Nat. Chem.}\ }\textbf {\bibinfo {volume} {8}},\ \bibinfo {pages} {237 } (\bibinfo {year} {2016})}\BibitemShut {NoStop}%
\bibitem [{\citenamefont {Ren}\ \emph {et~al.}(2019)\citenamefont {Ren}, \citenamefont {Wang}, \citenamefont {Skitnevskaya}, \citenamefont {Trofimov}, \citenamefont {Gokhberg},\ and\ \citenamefont {Dorn}}]{Ren:2018}%
  \BibitemOpen
  \bibfield  {author} {\bibinfo {author} {\bibfnamefont {X.}~\bibnamefont {Ren}}, \bibinfo {author} {\bibfnamefont {E.}~\bibnamefont {Wang}}, \bibinfo {author} {\bibfnamefont {A.~D.}\ \bibnamefont {Skitnevskaya}}, \bibinfo {author} {\bibfnamefont {A.~B.}\ \bibnamefont {Trofimov}}, \bibinfo {author} {\bibfnamefont {K.}~\bibnamefont {Gokhberg}},\ and\ \bibinfo {author} {\bibfnamefont {A.}~\bibnamefont {Dorn}},\ }\bibfield  {title} {\bibinfo {title} {Experimental evidence for ultrafast intermolecular relaxation processes in hydrated biomolecules},\ }\href@noop {} {\bibfield  {journal} {\bibinfo  {journal} {Nat. Phys.}\ }\textbf {\bibinfo {volume} {1}},\  (\bibinfo {year} {2019})}\BibitemShut {NoStop}%
\bibitem [{\citenamefont {Zhang}\ \emph {et~al.}(2022)\citenamefont {Zhang}, \citenamefont {Perry}, \citenamefont {Luu}, \citenamefont {Matselyukh},\ and\ \citenamefont {W\"orner}}]{Zhang:2022}%
  \BibitemOpen
  \bibfield  {author} {\bibinfo {author} {\bibfnamefont {P.}~\bibnamefont {Zhang}}, \bibinfo {author} {\bibfnamefont {C.}~\bibnamefont {Perry}}, \bibinfo {author} {\bibfnamefont {T.~T.}\ \bibnamefont {Luu}}, \bibinfo {author} {\bibfnamefont {D.}~\bibnamefont {Matselyukh}},\ and\ \bibinfo {author} {\bibfnamefont {H.~J.}\ \bibnamefont {W\"orner}},\ }\bibfield  {title} {\bibinfo {title} {Intermolecular coulombic decay in liquid water},\ }\href@noop {} {\bibfield  {journal} {\bibinfo  {journal} {Phys. Rev. Lett.}\ }\textbf {\bibinfo {volume} {128}},\ \bibinfo {pages} {133001} (\bibinfo {year} {2022})}\BibitemShut {NoStop}%
\bibitem [{\citenamefont {Ren}\ \emph {et~al.}(2023)\citenamefont {Ren}, \citenamefont {Wang}, \citenamefont {Zhou}, \citenamefont {Jia}, \citenamefont {Wang}, \citenamefont {Xue},\ and\ \citenamefont {Dorn}}]{Ren:2023}%
  \BibitemOpen
  \bibfield  {author} {\bibinfo {author} {\bibfnamefont {X.}~\bibnamefont {Ren}}, \bibinfo {author} {\bibfnamefont {E.}~\bibnamefont {Wang}}, \bibinfo {author} {\bibfnamefont {J.}~\bibnamefont {Zhou}}, \bibinfo {author} {\bibfnamefont {S.}~\bibnamefont {Jia}}, \bibinfo {author} {\bibfnamefont {X.}~\bibnamefont {Wang}}, \bibinfo {author} {\bibfnamefont {X.}~\bibnamefont {Xue}},\ and\ \bibinfo {author} {\bibfnamefont {A.}~\bibnamefont {Dorn}},\ }\bibfield  {title} {\bibinfo {title} {Ultrafast molecular dissociation induced by intermolecular coulombic decay in water clusters},\ }\href@noop {} {\bibfield  {journal} {\bibinfo  {journal} {Phys. Rev. A}\ }\textbf {\bibinfo {volume} {108}},\ \bibinfo {pages} {052814} (\bibinfo {year} {2023})}\BibitemShut {NoStop}%
\bibitem [{\citenamefont {Gopakumar}\ \emph {et~al.}(2023)\citenamefont {Gopakumar}, \citenamefont {Unger}, \citenamefont {Slav{\'\i}{\v{c}}ek}, \citenamefont {Hergenhahn}, \citenamefont {{\"O}hrwall}, \citenamefont {Malerz}, \citenamefont {C{\'e}olin}, \citenamefont {Trinter}, \citenamefont {Winter}, \citenamefont {Wilkinson} \emph {et~al.}}]{gopakumar2023radiation}%
  \BibitemOpen
  \bibfield  {author} {\bibinfo {author} {\bibfnamefont {G.}~\bibnamefont {Gopakumar}}, \bibinfo {author} {\bibfnamefont {I.}~\bibnamefont {Unger}}, \bibinfo {author} {\bibfnamefont {P.}~\bibnamefont {Slav{\'\i}{\v{c}}ek}}, \bibinfo {author} {\bibfnamefont {U.}~\bibnamefont {Hergenhahn}}, \bibinfo {author} {\bibfnamefont {G.}~\bibnamefont {{\"O}hrwall}}, \bibinfo {author} {\bibfnamefont {S.}~\bibnamefont {Malerz}}, \bibinfo {author} {\bibfnamefont {D.}~\bibnamefont {C{\'e}olin}}, \bibinfo {author} {\bibfnamefont {F.}~\bibnamefont {Trinter}}, \bibinfo {author} {\bibfnamefont {B.}~\bibnamefont {Winter}}, \bibinfo {author} {\bibfnamefont {I.}~\bibnamefont {Wilkinson}}, \emph {et~al.},\ }\bibfield  {title} {\bibinfo {title} {Radiation damage by extensive local water ionization from two-step electron-transfer-mediated decay of solvated ions},\ }\href@noop {} {\bibfield  {journal} {\bibinfo  {journal} {Nat. Chem.}\ }\textbf {\bibinfo {volume} {15}},\ \bibinfo {pages} {1408} (\bibinfo {year} {2023})}\BibitemShut
  {NoStop}%
\bibitem [{\citenamefont {Mudrich}\ \emph {et~al.}(2020)\citenamefont {Mudrich}, \citenamefont {LaForge}, \citenamefont {Ciavardini}, \citenamefont {O'Keeffe}, \citenamefont {Callegari}, \citenamefont {Coreno}, \citenamefont {Demidovich}, \citenamefont {Devetta}, \citenamefont {Di~Fraia}, \citenamefont {Drabbels} \emph {et~al.}}]{Mudrich:2020}%
  \BibitemOpen
  \bibfield  {author} {\bibinfo {author} {\bibfnamefont {M.}~\bibnamefont {Mudrich}}, \bibinfo {author} {\bibfnamefont {A.}~\bibnamefont {LaForge}}, \bibinfo {author} {\bibfnamefont {A.}~\bibnamefont {Ciavardini}}, \bibinfo {author} {\bibfnamefont {P.}~\bibnamefont {O'Keeffe}}, \bibinfo {author} {\bibfnamefont {C.}~\bibnamefont {Callegari}}, \bibinfo {author} {\bibfnamefont {M.}~\bibnamefont {Coreno}}, \bibinfo {author} {\bibfnamefont {A.}~\bibnamefont {Demidovich}}, \bibinfo {author} {\bibfnamefont {M.}~\bibnamefont {Devetta}}, \bibinfo {author} {\bibfnamefont {M.}~\bibnamefont {Di~Fraia}}, \bibinfo {author} {\bibfnamefont {M.}~\bibnamefont {Drabbels}}, \emph {et~al.},\ }\bibfield  {title} {\bibinfo {title} {Ultrafast relaxation of photoexcited superfluid he nanodroplets},\ }\href@noop {} {\bibfield  {journal} {\bibinfo  {journal} {Nat. Commun.}\ }\textbf {\bibinfo {volume} {11}} (\bibinfo {year} {2020})}\BibitemShut {NoStop}%
\bibitem [{\citenamefont {Toennies}\ and\ \citenamefont {Vilesov}(2004)}]{Toennies:2004}%
  \BibitemOpen
  \bibfield  {author} {\bibinfo {author} {\bibfnamefont {J.~P.}\ \bibnamefont {Toennies}}\ and\ \bibinfo {author} {\bibfnamefont {A.~F.}\ \bibnamefont {Vilesov}},\ }\bibfield  {title} {\bibinfo {title} {Superfluid helium droplets: A uniquely cold nanomatrix for molecules and molecular complexes},\ }\href@noop {} {\bibfield  {journal} {\bibinfo  {journal} {Angew. Chem., Int. Ed. Engl.}\ }\textbf {\bibinfo {volume} {43}},\ \bibinfo {pages} {2622} (\bibinfo {year} {2004})}\BibitemShut {NoStop}%
\bibitem [{\citenamefont {Mudrich}\ and\ \citenamefont {Stienkemeier}(2014)}]{Mudrich:2014}%
  \BibitemOpen
  \bibfield  {author} {\bibinfo {author} {\bibfnamefont {M.}~\bibnamefont {Mudrich}}\ and\ \bibinfo {author} {\bibfnamefont {F.}~\bibnamefont {Stienkemeier}},\ }\bibfield  {title} {\bibinfo {title} {Photoionisaton of pure and doped helium nanodroplets},\ }\href@noop {} {\bibfield  {journal} {\bibinfo  {journal} {Int. Rev. Phys. Chem.}\ }\textbf {\bibinfo {volume} {33}},\ \bibinfo {pages} {301} (\bibinfo {year} {2014})}\BibitemShut {NoStop}%
\bibitem [{\citenamefont {Slenczka}\ and\ \citenamefont {Toennies}(2022)}]{slenczka2022molecules}%
  \BibitemOpen
  \bibfield  {author} {\bibinfo {author} {\bibfnamefont {A.}~\bibnamefont {Slenczka}}\ and\ \bibinfo {author} {\bibfnamefont {J.~P.}\ \bibnamefont {Toennies}},\ }\href@noop {} {\emph {\bibinfo {title} {Molecules in Superfluid Helium Nanodroplets: Spectroscopy, Structure, and Dynamics}}}\ (\bibinfo  {publisher} {Springer Nature},\ \bibinfo {year} {2022})\BibitemShut {NoStop}%
\bibitem [{\citenamefont {Fr\"{o}chtenicht}\ \emph {et~al.}(1996)\citenamefont {Fr\"{o}chtenicht}, \citenamefont {Henne}, \citenamefont {Toennies}, \citenamefont {Ding}, \citenamefont {Fieber-Erdmann},\ and\ \citenamefont {Drewello}}]{Froechtenicht:1996}%
  \BibitemOpen
  \bibfield  {author} {\bibinfo {author} {\bibfnamefont {R.}~\bibnamefont {Fr\"{o}chtenicht}}, \bibinfo {author} {\bibfnamefont {U.}~\bibnamefont {Henne}}, \bibinfo {author} {\bibfnamefont {J.~P.}\ \bibnamefont {Toennies}}, \bibinfo {author} {\bibfnamefont {A.}~\bibnamefont {Ding}}, \bibinfo {author} {\bibfnamefont {M.}~\bibnamefont {Fieber-Erdmann}},\ and\ \bibinfo {author} {\bibfnamefont {T.}~\bibnamefont {Drewello}},\ }\bibfield  {title} {\bibinfo {title} {The photoionization of large pure and doped helium droplets},\ }\href@noop {} {\bibfield  {journal} {\bibinfo  {journal} {J. Chem. Phys.}\ }\textbf {\bibinfo {volume} {104}},\ \bibinfo {pages} {2548} (\bibinfo {year} {1996})}\BibitemShut {NoStop}%
\bibitem [{\citenamefont {Buchta}\ \emph {et~al.}(2013{\natexlab{a}})\citenamefont {Buchta}, \citenamefont {Krishnan}, \citenamefont {Brauer}, \citenamefont {Drabbels}, \citenamefont {O'Keeffe}, \citenamefont {Devetta}, \citenamefont {Di~Fraia}, \citenamefont {Callegari}, \citenamefont {Richter}, \citenamefont {Coreno}, \citenamefont {Prince}, \citenamefont {Stienkemeier}, \citenamefont {Ullrich}, \citenamefont {Moshammer},\ and\ \citenamefont {Mudrich}}]{BuchtaJCP:2013}%
  \BibitemOpen
  \bibfield  {author} {\bibinfo {author} {\bibfnamefont {D.}~\bibnamefont {Buchta}}, \bibinfo {author} {\bibfnamefont {S.~R.}\ \bibnamefont {Krishnan}}, \bibinfo {author} {\bibfnamefont {N.~B.}\ \bibnamefont {Brauer}}, \bibinfo {author} {\bibfnamefont {M.}~\bibnamefont {Drabbels}}, \bibinfo {author} {\bibfnamefont {P.}~\bibnamefont {O'Keeffe}}, \bibinfo {author} {\bibfnamefont {M.}~\bibnamefont {Devetta}}, \bibinfo {author} {\bibfnamefont {M.}~\bibnamefont {Di~Fraia}}, \bibinfo {author} {\bibfnamefont {C.}~\bibnamefont {Callegari}}, \bibinfo {author} {\bibfnamefont {R.}~\bibnamefont {Richter}}, \bibinfo {author} {\bibfnamefont {M.}~\bibnamefont {Coreno}}, \bibinfo {author} {\bibfnamefont {K.~C.}\ \bibnamefont {Prince}}, \bibinfo {author} {\bibfnamefont {F.}~\bibnamefont {Stienkemeier}}, \bibinfo {author} {\bibfnamefont {J.}~\bibnamefont {Ullrich}}, \bibinfo {author} {\bibfnamefont {R.}~\bibnamefont {Moshammer}},\ and\ \bibinfo {author} {\bibfnamefont {M.}~\bibnamefont {Mudrich}},\ }\bibfield  {title}
  {\bibinfo {title} {Extreme ultraviolet ionization of pure he nanodroplets: Mass-correlated photoelectron imaging, penning ionization, and electron energy-loss spectra},\ }\href@noop {} {\bibfield  {journal} {\bibinfo  {journal} {J. Chem. Phys.}\ }\textbf {\bibinfo {volume} {139}},\ \bibinfo {pages} {084301} (\bibinfo {year} {2013}{\natexlab{a}})}\BibitemShut {NoStop}%
\bibitem [{\citenamefont {LaForge}\ \emph {et~al.}(2016)\citenamefont {LaForge}, \citenamefont {Stumpf}, \citenamefont {Gokhberg}, \citenamefont {von Vangerow}, \citenamefont {Stienkemeier}, \citenamefont {Kryzhevoi}, \citenamefont {O'Keeffe}, \citenamefont {Ciavardini}, \citenamefont {Krishnan}, \citenamefont {Coreno}, \citenamefont {Prince}, \citenamefont {Richter}, \citenamefont {Moshammer}, \citenamefont {Pfeifer}, \citenamefont {Cederbaum},\ and\ \citenamefont {Mudrich}}]{LaForgePRL:2016}%
  \BibitemOpen
  \bibfield  {author} {\bibinfo {author} {\bibfnamefont {A.~C.}\ \bibnamefont {LaForge}}, \bibinfo {author} {\bibfnamefont {V.}~\bibnamefont {Stumpf}}, \bibinfo {author} {\bibfnamefont {K.}~\bibnamefont {Gokhberg}}, \bibinfo {author} {\bibfnamefont {J.}~\bibnamefont {von Vangerow}}, \bibinfo {author} {\bibfnamefont {F.}~\bibnamefont {Stienkemeier}}, \bibinfo {author} {\bibfnamefont {N.~V.}\ \bibnamefont {Kryzhevoi}}, \bibinfo {author} {\bibfnamefont {P.}~\bibnamefont {O'Keeffe}}, \bibinfo {author} {\bibfnamefont {A.}~\bibnamefont {Ciavardini}}, \bibinfo {author} {\bibfnamefont {S.~R.}\ \bibnamefont {Krishnan}}, \bibinfo {author} {\bibfnamefont {M.}~\bibnamefont {Coreno}}, \bibinfo {author} {\bibfnamefont {K.~C.}\ \bibnamefont {Prince}}, \bibinfo {author} {\bibfnamefont {R.}~\bibnamefont {Richter}}, \bibinfo {author} {\bibfnamefont {R.}~\bibnamefont {Moshammer}}, \bibinfo {author} {\bibfnamefont {T.}~\bibnamefont {Pfeifer}}, \bibinfo {author} {\bibfnamefont {L.~S.}\ \bibnamefont {Cederbaum}},\ and\ \bibinfo
  {author} {\bibfnamefont {M.}~\bibnamefont {Mudrich}},\ }\bibfield  {title} {\bibinfo {title} {Enhanced ionization of embedded clusters by electron-transfer-mediated decay in helium nanodroplets},\ }\href@noop {} {\bibfield  {journal} {\bibinfo  {journal} {Phys. Rev. Lett.}\ }\textbf {\bibinfo {volume} {116}},\ \bibinfo {pages} {203001} (\bibinfo {year} {2016})}\BibitemShut {NoStop}%
\bibitem [{\citenamefont {Shcherbinin}\ \emph {et~al.}(2017)\citenamefont {Shcherbinin}, \citenamefont {LaForge}, \citenamefont {Sharma}, \citenamefont {Devetta}, \citenamefont {Richter}, \citenamefont {Moshammer}, \citenamefont {Pfeifer},\ and\ \citenamefont {Mudrich}}]{Shcherbinin:2017}%
  \BibitemOpen
  \bibfield  {author} {\bibinfo {author} {\bibfnamefont {M.}~\bibnamefont {Shcherbinin}}, \bibinfo {author} {\bibfnamefont {A.~C.}\ \bibnamefont {LaForge}}, \bibinfo {author} {\bibfnamefont {V.}~\bibnamefont {Sharma}}, \bibinfo {author} {\bibfnamefont {M.}~\bibnamefont {Devetta}}, \bibinfo {author} {\bibfnamefont {R.}~\bibnamefont {Richter}}, \bibinfo {author} {\bibfnamefont {R.}~\bibnamefont {Moshammer}}, \bibinfo {author} {\bibfnamefont {T.}~\bibnamefont {Pfeifer}},\ and\ \bibinfo {author} {\bibfnamefont {M.}~\bibnamefont {Mudrich}},\ }\bibfield  {title} {\bibinfo {title} {Interatomic coulombic decay in helium nanodroplets},\ }\href@noop {} {\bibfield  {journal} {\bibinfo  {journal} {Phys. Rev. A}\ }\textbf {\bibinfo {volume} {96}},\ \bibinfo {pages} {013407} (\bibinfo {year} {2017})}\BibitemShut {NoStop}%
\bibitem [{\citenamefont {Shcherbinin}\ \emph {et~al.}(2018)\citenamefont {Shcherbinin}, \citenamefont {LaForge}, \citenamefont {Hanif}, \citenamefont {Richter},\ and\ \citenamefont {Mudrich}}]{Shcherbinin:2018}%
  \BibitemOpen
  \bibfield  {author} {\bibinfo {author} {\bibfnamefont {M.}~\bibnamefont {Shcherbinin}}, \bibinfo {author} {\bibfnamefont {A.~C.}\ \bibnamefont {LaForge}}, \bibinfo {author} {\bibfnamefont {M.}~\bibnamefont {Hanif}}, \bibinfo {author} {\bibfnamefont {R.}~\bibnamefont {Richter}},\ and\ \bibinfo {author} {\bibfnamefont {M.}~\bibnamefont {Mudrich}},\ }\bibfield  {title} {\bibinfo {title} {Penning ionization of acene molecules by helium nanodroplets},\ }\href@noop {} {\bibfield  {journal} {\bibinfo  {journal} {J. Phys. Chem. A}\ }\textbf {\bibinfo {volume} {122}},\ \bibinfo {pages} {1855} (\bibinfo {year} {2018})}\BibitemShut {NoStop}%
\bibitem [{\citenamefont {Wiegandt}\ \emph {et~al.}(2019)\citenamefont {Wiegandt}, \citenamefont {Trinter}, \citenamefont {Henrichs}, \citenamefont {Metz}, \citenamefont {Pitzer}, \citenamefont {Waitz}, \citenamefont {Al~Maalouf}, \citenamefont {Janke}, \citenamefont {Rist}, \citenamefont {Wechselberger}, \citenamefont {Miteva}, \citenamefont {Kazandjian}, \citenamefont {Sch\"offler}, \citenamefont {Sisourat}, \citenamefont {Jahnke},\ and\ \citenamefont {D\"orner}}]{wiegandt:2019}%
  \BibitemOpen
  \bibfield  {author} {\bibinfo {author} {\bibfnamefont {F.}~\bibnamefont {Wiegandt}}, \bibinfo {author} {\bibfnamefont {F.}~\bibnamefont {Trinter}}, \bibinfo {author} {\bibfnamefont {K.}~\bibnamefont {Henrichs}}, \bibinfo {author} {\bibfnamefont {D.}~\bibnamefont {Metz}}, \bibinfo {author} {\bibfnamefont {M.}~\bibnamefont {Pitzer}}, \bibinfo {author} {\bibfnamefont {M.}~\bibnamefont {Waitz}}, \bibinfo {author} {\bibfnamefont {E.~J.}\ \bibnamefont {Al~Maalouf}}, \bibinfo {author} {\bibfnamefont {C.}~\bibnamefont {Janke}}, \bibinfo {author} {\bibfnamefont {J.}~\bibnamefont {Rist}}, \bibinfo {author} {\bibfnamefont {N.}~\bibnamefont {Wechselberger}}, \bibinfo {author} {\bibfnamefont {T.}~\bibnamefont {Miteva}}, \bibinfo {author} {\bibfnamefont {S.}~\bibnamefont {Kazandjian}}, \bibinfo {author} {\bibfnamefont {M.}~\bibnamefont {Sch\"offler}}, \bibinfo {author} {\bibfnamefont {N.}~\bibnamefont {Sisourat}}, \bibinfo {author} {\bibfnamefont {T.}~\bibnamefont {Jahnke}},\ and\ \bibinfo {author} {\bibfnamefont
  {R.}~\bibnamefont {D\"orner}},\ }\bibfield  {title} {\bibinfo {title} {Direct observation of interatomic coulombic decay and subsequent ion-atom scattering in helium nanodroplets},\ }\href@noop {} {\bibfield  {journal} {\bibinfo  {journal} {Phys. Rev. A}\ }\textbf {\bibinfo {volume} {100}},\ \bibinfo {pages} {022707} (\bibinfo {year} {2019})}\BibitemShut {NoStop}%
\bibitem [{\citenamefont {Ben~Ltaief}\ \emph {et~al.}(2019)\citenamefont {Ben~Ltaief}, \citenamefont {Shcherbinin}, \citenamefont {Mandal}, \citenamefont {Krishnan}, \citenamefont {LaForge}, \citenamefont {Richter}, \citenamefont {Turchini}, \citenamefont {Zema}, \citenamefont {Pfeifer}, \citenamefont {Fasshauer} \emph {et~al.}}]{Ltaief:2019}%
  \BibitemOpen
  \bibfield  {author} {\bibinfo {author} {\bibfnamefont {L.}~\bibnamefont {Ben~Ltaief}}, \bibinfo {author} {\bibfnamefont {M.}~\bibnamefont {Shcherbinin}}, \bibinfo {author} {\bibfnamefont {S.}~\bibnamefont {Mandal}}, \bibinfo {author} {\bibfnamefont {S.}~\bibnamefont {Krishnan}}, \bibinfo {author} {\bibfnamefont {A.}~\bibnamefont {LaForge}}, \bibinfo {author} {\bibfnamefont {R.}~\bibnamefont {Richter}}, \bibinfo {author} {\bibfnamefont {S.}~\bibnamefont {Turchini}}, \bibinfo {author} {\bibfnamefont {N.}~\bibnamefont {Zema}}, \bibinfo {author} {\bibfnamefont {T.}~\bibnamefont {Pfeifer}}, \bibinfo {author} {\bibfnamefont {E.}~\bibnamefont {Fasshauer}}, \emph {et~al.},\ }\bibfield  {title} {\bibinfo {title} {Charge exchange dominates long-range interatomic coulombic decay of excited metal-doped helium nanodroplets},\ }\href@noop {} {\bibfield  {journal} {\bibinfo  {journal} {J. Phys. Chem. Lett.}\ }\textbf {\bibinfo {volume} {10}},\ \bibinfo {pages} {6904} (\bibinfo {year} {2019})}\BibitemShut {NoStop}%
\bibitem [{\citenamefont {Ltaief}\ \emph {et~al.}(2020)\citenamefont {Ltaief}, \citenamefont {Shcherbinin}, \citenamefont {Krishnan}, \citenamefont {Richter}, \citenamefont {Pfeifer}, \citenamefont {Bauer}, \citenamefont {Ghosh}, \citenamefont {Mudrich}, \citenamefont {Gokhberg}, \citenamefont {Laforge} \emph {et~al.}}]{Ltaief:2020}%
  \BibitemOpen
  \bibfield  {author} {\bibinfo {author} {\bibfnamefont {L.~B.}\ \bibnamefont {Ltaief}}, \bibinfo {author} {\bibfnamefont {M.}~\bibnamefont {Shcherbinin}}, \bibinfo {author} {\bibfnamefont {S.}~\bibnamefont {Krishnan}}, \bibinfo {author} {\bibfnamefont {R.}~\bibnamefont {Richter}}, \bibinfo {author} {\bibfnamefont {T.}~\bibnamefont {Pfeifer}}, \bibinfo {author} {\bibfnamefont {M.}~\bibnamefont {Bauer}}, \bibinfo {author} {\bibfnamefont {A.}~\bibnamefont {Ghosh}}, \bibinfo {author} {\bibfnamefont {M.}~\bibnamefont {Mudrich}}, \bibinfo {author} {\bibfnamefont {K.}~\bibnamefont {Gokhberg}}, \bibinfo {author} {\bibfnamefont {A.}~\bibnamefont {Laforge}}, \emph {et~al.},\ }\bibfield  {title} {\bibinfo {title} {Electron transfer mediated decay of alkali dimers attached to he nanodroplets},\ }\href@noop {} {\bibfield  {journal} {\bibinfo  {journal} {Phys. Chem. Chem. Phys.}\ } (\bibinfo {year} {2020})}\BibitemShut {NoStop}%
\bibitem [{\citenamefont {Ben~Ltaief}\ \emph {et~al.}(2023)\citenamefont {Ben~Ltaief}, \citenamefont {Sishodia}, \citenamefont {Mandal}, \citenamefont {De}, \citenamefont {Krishnan}, \citenamefont {Medina}, \citenamefont {Pal}, \citenamefont {Richter}, \citenamefont {Fennel},\ and\ \citenamefont {Mudrich}}]{ltaief:2023}%
  \BibitemOpen
  \bibfield  {author} {\bibinfo {author} {\bibfnamefont {L.}~\bibnamefont {Ben~Ltaief}}, \bibinfo {author} {\bibfnamefont {K.}~\bibnamefont {Sishodia}}, \bibinfo {author} {\bibfnamefont {S.}~\bibnamefont {Mandal}}, \bibinfo {author} {\bibfnamefont {S.}~\bibnamefont {De}}, \bibinfo {author} {\bibfnamefont {S.~R.}\ \bibnamefont {Krishnan}}, \bibinfo {author} {\bibfnamefont {C.}~\bibnamefont {Medina}}, \bibinfo {author} {\bibfnamefont {N.}~\bibnamefont {Pal}}, \bibinfo {author} {\bibfnamefont {R.}~\bibnamefont {Richter}}, \bibinfo {author} {\bibfnamefont {T.}~\bibnamefont {Fennel}},\ and\ \bibinfo {author} {\bibfnamefont {M.}~\bibnamefont {Mudrich}},\ }\bibfield  {title} {\bibinfo {title} {Efficient indirect interatomic coulombic decay induced by photoelectron impact excitation in large pure helium nanodroplets},\ }\href@noop {} {\bibfield  {journal} {\bibinfo  {journal} {Phys. Rev. Lett.}\ }\textbf {\bibinfo {volume} {131}},\ \bibinfo {pages} {023001} (\bibinfo {year} {2023})}\BibitemShut {NoStop}%
\bibitem [{\citenamefont {Asmussen}\ \emph {et~al.}(2023{\natexlab{a}})\citenamefont {Asmussen}, \citenamefont {Ben~Ltaief}, \citenamefont {Sishodia}, \citenamefont {Abid}, \citenamefont {Bastian}, \citenamefont {Krishnan}, \citenamefont {Pedersen},\ and\ \citenamefont {Mudrich}}]{asmussen2023dopant}%
  \BibitemOpen
  \bibfield  {author} {\bibinfo {author} {\bibfnamefont {J.~D.}\ \bibnamefont {Asmussen}}, \bibinfo {author} {\bibfnamefont {L.}~\bibnamefont {Ben~Ltaief}}, \bibinfo {author} {\bibfnamefont {K.}~\bibnamefont {Sishodia}}, \bibinfo {author} {\bibfnamefont {A.~R.}\ \bibnamefont {Abid}}, \bibinfo {author} {\bibfnamefont {B.}~\bibnamefont {Bastian}}, \bibinfo {author} {\bibfnamefont {S.}~\bibnamefont {Krishnan}}, \bibinfo {author} {\bibfnamefont {H.~B.}\ \bibnamefont {Pedersen}},\ and\ \bibinfo {author} {\bibfnamefont {M.}~\bibnamefont {Mudrich}},\ }\bibfield  {title} {\bibinfo {title} {Dopant ionization and efficiency of ion and electron ejection from helium nanodroplets},\ }\href@noop {} {\bibfield  {journal} {\bibinfo  {journal} {J. Chem. Phys.}\ }\textbf {\bibinfo {volume} {159}} (\bibinfo {year} {2023}{\natexlab{a}})}\BibitemShut {NoStop}%
\bibitem [{\citenamefont {Asmussen}\ \emph {et~al.}(2023{\natexlab{b}})\citenamefont {Asmussen}, \citenamefont {Abid}, \citenamefont {Sundaralingam}, \citenamefont {Bastian}, \citenamefont {Sishodia}, \citenamefont {De}, \citenamefont {Ltaief}, \citenamefont {Krishnan}, \citenamefont {Pedersen},\ and\ \citenamefont {Mudrich}}]{asmussen2023secondary}%
  \BibitemOpen
  \bibfield  {author} {\bibinfo {author} {\bibfnamefont {J.~D.}\ \bibnamefont {Asmussen}}, \bibinfo {author} {\bibfnamefont {A.~R.}\ \bibnamefont {Abid}}, \bibinfo {author} {\bibfnamefont {A.}~\bibnamefont {Sundaralingam}}, \bibinfo {author} {\bibfnamefont {B.}~\bibnamefont {Bastian}}, \bibinfo {author} {\bibfnamefont {K.}~\bibnamefont {Sishodia}}, \bibinfo {author} {\bibfnamefont {S.}~\bibnamefont {De}}, \bibinfo {author} {\bibfnamefont {L.~B.}\ \bibnamefont {Ltaief}}, \bibinfo {author} {\bibfnamefont {S.}~\bibnamefont {Krishnan}}, \bibinfo {author} {\bibfnamefont {H.~B.}\ \bibnamefont {Pedersen}},\ and\ \bibinfo {author} {\bibfnamefont {M.}~\bibnamefont {Mudrich}},\ }\bibfield  {title} {\bibinfo {title} {Secondary ionization of pyrimidine nucleobases and their microhydrated derivatives in helium nanodroplets},\ }\href@noop {} {\bibfield  {journal} {\bibinfo  {journal} {Phys. Chem. Chem. Phys.}\ }\textbf {\bibinfo {volume} {25}},\ \bibinfo {pages} {24819} (\bibinfo {year} {2023}{\natexlab{b}})}\BibitemShut
  {NoStop}%
\bibitem [{\citenamefont {Ben~Ltaief}\ \emph {et~al.}(2024)\citenamefont {Ben~Ltaief}, \citenamefont {Sishodia}, \citenamefont {Richter}, \citenamefont {Bastian}, \citenamefont {Asmussen}, \citenamefont {Mandal}, \citenamefont {Pal}, \citenamefont {Medina}, \citenamefont {Krishnan}, \citenamefont {von Haeften},\ and\ \citenamefont {Mudrich}}]{ltaiefPRR:2024}%
  \BibitemOpen
  \bibfield  {author} {\bibinfo {author} {\bibfnamefont {L.}~\bibnamefont {Ben~Ltaief}}, \bibinfo {author} {\bibfnamefont {K.}~\bibnamefont {Sishodia}}, \bibinfo {author} {\bibfnamefont {R.}~\bibnamefont {Richter}}, \bibinfo {author} {\bibfnamefont {B.}~\bibnamefont {Bastian}}, \bibinfo {author} {\bibfnamefont {J.~D.}\ \bibnamefont {Asmussen}}, \bibinfo {author} {\bibfnamefont {S.}~\bibnamefont {Mandal}}, \bibinfo {author} {\bibfnamefont {N.}~\bibnamefont {Pal}}, \bibinfo {author} {\bibfnamefont {C.}~\bibnamefont {Medina}}, \bibinfo {author} {\bibfnamefont {S.~R.}\ \bibnamefont {Krishnan}}, \bibinfo {author} {\bibfnamefont {K.}~\bibnamefont {von Haeften}},\ and\ \bibinfo {author} {\bibfnamefont {M.}~\bibnamefont {Mudrich}},\ }\bibfield  {title} {\bibinfo {title} {Spectroscopically resolved resonant interatomic coulombic decay in photoexcited large he nanodroplets},\ }\href@noop {} {\bibfield  {journal} {\bibinfo  {journal} {Phys. Rev. Res.}\ }\textbf {\bibinfo {volume} {6}},\ \bibinfo {pages} {013019}
  (\bibinfo {year} {2024})}\BibitemShut {NoStop}%
\bibitem [{\citenamefont {Bastian}\ \emph {et~al.}(2024)\citenamefont {Bastian}, \citenamefont {Asmussen}, \citenamefont {Ltaief}, \citenamefont {Pedersen}, \citenamefont {Sishodia}, \citenamefont {De}, \citenamefont {Krishnan}, \citenamefont {Medina}, \citenamefont {Pal}, \citenamefont {Richter} \emph {et~al.}}]{bastian2024observation}%
  \BibitemOpen
  \bibfield  {author} {\bibinfo {author} {\bibfnamefont {B.}~\bibnamefont {Bastian}}, \bibinfo {author} {\bibfnamefont {J.}~\bibnamefont {Asmussen}}, \bibinfo {author} {\bibfnamefont {L.~B.}\ \bibnamefont {Ltaief}}, \bibinfo {author} {\bibfnamefont {H.}~\bibnamefont {Pedersen}}, \bibinfo {author} {\bibfnamefont {K.}~\bibnamefont {Sishodia}}, \bibinfo {author} {\bibfnamefont {S.}~\bibnamefont {De}}, \bibinfo {author} {\bibfnamefont {S.}~\bibnamefont {Krishnan}}, \bibinfo {author} {\bibfnamefont {C.}~\bibnamefont {Medina}}, \bibinfo {author} {\bibfnamefont {N.}~\bibnamefont {Pal}}, \bibinfo {author} {\bibfnamefont {R.}~\bibnamefont {Richter}}, \emph {et~al.},\ }\bibfield  {title} {\bibinfo {title} {Observation of interatomic coulombic decay induced by double excitation of helium in nanodroplets},\ }\href@noop {} {\bibfield  {journal} {\bibinfo  {journal} {Phys. Rev. Lett.}\ }\textbf {\bibinfo {volume} {132}},\ \bibinfo {pages} {233001} (\bibinfo {year} {2024})}\BibitemShut {NoStop}%
\bibitem [{\citenamefont {Jurkovi{\v{c}}ov{\'a}}\ \emph {et~al.}(2024)\citenamefont {Jurkovi{\v{c}}ov{\'a}}, \citenamefont {Ben~Ltaief}, \citenamefont {Hult~Roos}, \citenamefont {Hort}, \citenamefont {Finke}, \citenamefont {Albrecht}, \citenamefont {Hoque}, \citenamefont {Klime{\v{s}}ov{\'a}}, \citenamefont {Sundaralingam}, \citenamefont {Antipenkov} \emph {et~al.}}]{jurkovivcova:2024}%
  \BibitemOpen
  \bibfield  {author} {\bibinfo {author} {\bibfnamefont {L.}~\bibnamefont {Jurkovi{\v{c}}ov{\'a}}}, \bibinfo {author} {\bibfnamefont {L.}~\bibnamefont {Ben~Ltaief}}, \bibinfo {author} {\bibfnamefont {A.}~\bibnamefont {Hult~Roos}}, \bibinfo {author} {\bibfnamefont {O.}~\bibnamefont {Hort}}, \bibinfo {author} {\bibfnamefont {O.}~\bibnamefont {Finke}}, \bibinfo {author} {\bibfnamefont {M.}~\bibnamefont {Albrecht}}, \bibinfo {author} {\bibfnamefont {Z.}~\bibnamefont {Hoque}}, \bibinfo {author} {\bibfnamefont {E.}~\bibnamefont {Klime{\v{s}}ov{\'a}}}, \bibinfo {author} {\bibfnamefont {A.}~\bibnamefont {Sundaralingam}}, \bibinfo {author} {\bibfnamefont {R.}~\bibnamefont {Antipenkov}}, \emph {et~al.},\ }\bibfield  {title} {\bibinfo {title} {Bright continuously tunable vacuum ultraviolet source for ultrafast spectroscopy},\ }\href@noop {} {\bibfield  {journal} {\bibinfo  {journal} {Communications Physics}\ }\textbf {\bibinfo {volume} {7}},\ \bibinfo {pages} {26} (\bibinfo {year} {2024})}\BibitemShut {NoStop}%
\bibitem [{\citenamefont {Ovcharenko}\ \emph {et~al.}(2014)\citenamefont {Ovcharenko}, \citenamefont {Lyamayev}, \citenamefont {Katzy}, \citenamefont {Devetta}, \citenamefont {LaForge}, \citenamefont {O'Keeffe}, \citenamefont {Plekan}, \citenamefont {Finetti}, \citenamefont {Di~Fraia}, \citenamefont {Mudrich}, \citenamefont {Krikunova}, \citenamefont {Piseri}, \citenamefont {Coreno}, \citenamefont {Brauer}, \citenamefont {Mazza}, \citenamefont {Stranges}, \citenamefont {Grazioli}, \citenamefont {Richter}, \citenamefont {Prince}, \citenamefont {Drabbels}, \citenamefont {Callegari}, \citenamefont {Stienkemeier},\ and\ \citenamefont {M\"oller}}]{Ovcharenko:2014}%
  \BibitemOpen
  \bibfield  {author} {\bibinfo {author} {\bibfnamefont {Y.}~\bibnamefont {Ovcharenko}}, \bibinfo {author} {\bibfnamefont {V.}~\bibnamefont {Lyamayev}}, \bibinfo {author} {\bibfnamefont {R.}~\bibnamefont {Katzy}}, \bibinfo {author} {\bibfnamefont {M.}~\bibnamefont {Devetta}}, \bibinfo {author} {\bibfnamefont {A.}~\bibnamefont {LaForge}}, \bibinfo {author} {\bibfnamefont {P.}~\bibnamefont {O'Keeffe}}, \bibinfo {author} {\bibfnamefont {O.}~\bibnamefont {Plekan}}, \bibinfo {author} {\bibfnamefont {P.}~\bibnamefont {Finetti}}, \bibinfo {author} {\bibfnamefont {M.}~\bibnamefont {Di~Fraia}}, \bibinfo {author} {\bibfnamefont {M.}~\bibnamefont {Mudrich}}, \bibinfo {author} {\bibfnamefont {M.}~\bibnamefont {Krikunova}}, \bibinfo {author} {\bibfnamefont {P.}~\bibnamefont {Piseri}}, \bibinfo {author} {\bibfnamefont {M.}~\bibnamefont {Coreno}}, \bibinfo {author} {\bibfnamefont {N.~B.}\ \bibnamefont {Brauer}}, \bibinfo {author} {\bibfnamefont {T.}~\bibnamefont {Mazza}}, \bibinfo {author} {\bibfnamefont {S.}~\bibnamefont
  {Stranges}}, \bibinfo {author} {\bibfnamefont {C.}~\bibnamefont {Grazioli}}, \bibinfo {author} {\bibfnamefont {R.}~\bibnamefont {Richter}}, \bibinfo {author} {\bibfnamefont {K.~C.}\ \bibnamefont {Prince}}, \bibinfo {author} {\bibfnamefont {M.}~\bibnamefont {Drabbels}}, \bibinfo {author} {\bibfnamefont {C.}~\bibnamefont {Callegari}}, \bibinfo {author} {\bibfnamefont {F.}~\bibnamefont {Stienkemeier}},\ and\ \bibinfo {author} {\bibfnamefont {T.}~\bibnamefont {M\"oller}},\ }\bibfield  {title} {\bibinfo {title} {Novel collective autoionization process observed in electron spectra of he clusters},\ }\href@noop {} {\bibfield  {journal} {\bibinfo  {journal} {Phys. Rev. Lett.}\ }\textbf {\bibinfo {volume} {112}},\ \bibinfo {pages} {073401} (\bibinfo {year} {2014})}\BibitemShut {NoStop}%
\bibitem [{\citenamefont {LaForge}\ \emph {et~al.}(2014)\citenamefont {LaForge}, \citenamefont {Drabbels}, \citenamefont {Brauer}, \citenamefont {Coreno}, \citenamefont {Devetta}, \citenamefont {Di~Fraia}, \citenamefont {Finetti}, \citenamefont {Grazioli}, \citenamefont {Katzy}, \citenamefont {Lyamayev}, \citenamefont {Mazza}, \citenamefont {Mudrich}, \citenamefont {O'Keeffe}, \citenamefont {Ovcharenko}, \citenamefont {Piseri}, \citenamefont {Plekan}, \citenamefont {Prince}, \citenamefont {Richter}, \citenamefont {Stranges}, \citenamefont {Callegari}, \citenamefont {Moller},\ and\ \citenamefont {Stienkemeier}}]{LaForge:2014}%
  \BibitemOpen
  \bibfield  {author} {\bibinfo {author} {\bibfnamefont {A.~C.}\ \bibnamefont {LaForge}}, \bibinfo {author} {\bibfnamefont {M.}~\bibnamefont {Drabbels}}, \bibinfo {author} {\bibfnamefont {N.~B.}\ \bibnamefont {Brauer}}, \bibinfo {author} {\bibfnamefont {M.}~\bibnamefont {Coreno}}, \bibinfo {author} {\bibfnamefont {M.}~\bibnamefont {Devetta}}, \bibinfo {author} {\bibfnamefont {M.}~\bibnamefont {Di~Fraia}}, \bibinfo {author} {\bibfnamefont {P.}~\bibnamefont {Finetti}}, \bibinfo {author} {\bibfnamefont {C.}~\bibnamefont {Grazioli}}, \bibinfo {author} {\bibfnamefont {R.}~\bibnamefont {Katzy}}, \bibinfo {author} {\bibfnamefont {V.}~\bibnamefont {Lyamayev}}, \bibinfo {author} {\bibfnamefont {T.}~\bibnamefont {Mazza}}, \bibinfo {author} {\bibfnamefont {M.}~\bibnamefont {Mudrich}}, \bibinfo {author} {\bibfnamefont {P.}~\bibnamefont {O'Keeffe}}, \bibinfo {author} {\bibfnamefont {Y.}~\bibnamefont {Ovcharenko}}, \bibinfo {author} {\bibfnamefont {P.}~\bibnamefont {Piseri}}, \bibinfo {author} {\bibfnamefont
  {O.}~\bibnamefont {Plekan}}, \bibinfo {author} {\bibfnamefont {K.~C.}\ \bibnamefont {Prince}}, \bibinfo {author} {\bibfnamefont {R.}~\bibnamefont {Richter}}, \bibinfo {author} {\bibfnamefont {S.}~\bibnamefont {Stranges}}, \bibinfo {author} {\bibfnamefont {C.}~\bibnamefont {Callegari}}, \bibinfo {author} {\bibfnamefont {T.}~\bibnamefont {Moller}},\ and\ \bibinfo {author} {\bibfnamefont {F.}~\bibnamefont {Stienkemeier}},\ }\bibfield  {title} {\bibinfo {title} {Collective autoionization in multiply-excited systems: A novel ionization process observed in helium nanodroplets},\ }\href@noop {} {\bibfield  {journal} {\bibinfo  {journal} {Sci. Rep.}\ }\textbf {\bibinfo {volume} {4}},\ \bibinfo {pages} {3621} (\bibinfo {year} {2014})}\BibitemShut {NoStop}%
\bibitem [{\citenamefont {Ovcharenko}\ \emph {et~al.}(2020)\citenamefont {Ovcharenko}, \citenamefont {LaForge}, \citenamefont {Langbehn}, \citenamefont {Plekan}, \citenamefont {Cucini}, \citenamefont {Finetti}, \citenamefont {O’Keeffe}, \citenamefont {Iablonskyi}, \citenamefont {Nishiyama}, \citenamefont {Ueda} \emph {et~al.}}]{Ovcharenko:2020}%
  \BibitemOpen
  \bibfield  {author} {\bibinfo {author} {\bibfnamefont {Y.}~\bibnamefont {Ovcharenko}}, \bibinfo {author} {\bibfnamefont {A.}~\bibnamefont {LaForge}}, \bibinfo {author} {\bibfnamefont {B.}~\bibnamefont {Langbehn}}, \bibinfo {author} {\bibfnamefont {O.}~\bibnamefont {Plekan}}, \bibinfo {author} {\bibfnamefont {R.}~\bibnamefont {Cucini}}, \bibinfo {author} {\bibfnamefont {P.}~\bibnamefont {Finetti}}, \bibinfo {author} {\bibfnamefont {P.}~\bibnamefont {O’Keeffe}}, \bibinfo {author} {\bibfnamefont {D.}~\bibnamefont {Iablonskyi}}, \bibinfo {author} {\bibfnamefont {T.}~\bibnamefont {Nishiyama}}, \bibinfo {author} {\bibfnamefont {K.}~\bibnamefont {Ueda}}, \emph {et~al.},\ }\bibfield  {title} {\bibinfo {title} {Autoionization dynamics of helium nanodroplets resonantly excited by intense xuv laser pulses},\ }\href@noop {} {\bibfield  {journal} {\bibinfo  {journal} {New J. Phys.}\ }\textbf {\bibinfo {volume} {22}},\ \bibinfo {pages} {083043} (\bibinfo {year} {2020})}\BibitemShut {NoStop}%
\bibitem [{\citenamefont {LaForge}\ \emph {et~al.}(2021)\citenamefont {LaForge}, \citenamefont {Michiels}, \citenamefont {Ovcharenko}, \citenamefont {Ngai}, \citenamefont {Escart\'{\i}n}, \citenamefont {Berrah}, \citenamefont {Callegari}, \citenamefont {Clark}, \citenamefont {Coreno}, \citenamefont {Cucini}, \citenamefont {Di~Fraia}, \citenamefont {Drabbels}, \citenamefont {Fasshauer}, \citenamefont {Finetti}, \citenamefont {Giannessi}, \citenamefont {Grazioli}, \citenamefont {Iablonskyi}, \citenamefont {Langbehn}, \citenamefont {Nishiyama}, \citenamefont {Oliver}, \citenamefont {Piseri}, \citenamefont {Plekan}, \citenamefont {Prince}, \citenamefont {Rupp}, \citenamefont {Stranges}, \citenamefont {Ueda}, \citenamefont {Sisourat}, \citenamefont {Eloranta}, \citenamefont {Pi}, \citenamefont {Barranco}, \citenamefont {Stienkemeier}, \citenamefont {M\"oller},\ and\ \citenamefont {Mudrich}}]{LaForge:2021}%
  \BibitemOpen
  \bibfield  {author} {\bibinfo {author} {\bibfnamefont {A.~C.}\ \bibnamefont {LaForge}}, \bibinfo {author} {\bibfnamefont {R.}~\bibnamefont {Michiels}}, \bibinfo {author} {\bibfnamefont {Y.}~\bibnamefont {Ovcharenko}}, \bibinfo {author} {\bibfnamefont {A.}~\bibnamefont {Ngai}}, \bibinfo {author} {\bibfnamefont {J.~M.}\ \bibnamefont {Escart\'{\i}n}}, \bibinfo {author} {\bibfnamefont {N.}~\bibnamefont {Berrah}}, \bibinfo {author} {\bibfnamefont {C.}~\bibnamefont {Callegari}}, \bibinfo {author} {\bibfnamefont {A.}~\bibnamefont {Clark}}, \bibinfo {author} {\bibfnamefont {M.}~\bibnamefont {Coreno}}, \bibinfo {author} {\bibfnamefont {R.}~\bibnamefont {Cucini}}, \bibinfo {author} {\bibfnamefont {M.}~\bibnamefont {Di~Fraia}}, \bibinfo {author} {\bibfnamefont {M.}~\bibnamefont {Drabbels}}, \bibinfo {author} {\bibfnamefont {E.}~\bibnamefont {Fasshauer}}, \bibinfo {author} {\bibfnamefont {P.}~\bibnamefont {Finetti}}, \bibinfo {author} {\bibfnamefont {L.}~\bibnamefont {Giannessi}}, \bibinfo {author} {\bibfnamefont
  {C.}~\bibnamefont {Grazioli}}, \bibinfo {author} {\bibfnamefont {D.}~\bibnamefont {Iablonskyi}}, \bibinfo {author} {\bibfnamefont {B.}~\bibnamefont {Langbehn}}, \bibinfo {author} {\bibfnamefont {T.}~\bibnamefont {Nishiyama}}, \bibinfo {author} {\bibfnamefont {V.}~\bibnamefont {Oliver}}, \bibinfo {author} {\bibfnamefont {P.}~\bibnamefont {Piseri}}, \bibinfo {author} {\bibfnamefont {O.}~\bibnamefont {Plekan}}, \bibinfo {author} {\bibfnamefont {K.~C.}\ \bibnamefont {Prince}}, \bibinfo {author} {\bibfnamefont {D.}~\bibnamefont {Rupp}}, \bibinfo {author} {\bibfnamefont {S.}~\bibnamefont {Stranges}}, \bibinfo {author} {\bibfnamefont {K.}~\bibnamefont {Ueda}}, \bibinfo {author} {\bibfnamefont {N.}~\bibnamefont {Sisourat}}, \bibinfo {author} {\bibfnamefont {J.}~\bibnamefont {Eloranta}}, \bibinfo {author} {\bibfnamefont {M.}~\bibnamefont {Pi}}, \bibinfo {author} {\bibfnamefont {M.}~\bibnamefont {Barranco}}, \bibinfo {author} {\bibfnamefont {F.}~\bibnamefont {Stienkemeier}}, \bibinfo {author} {\bibfnamefont
  {T.}~\bibnamefont {M\"oller}},\ and\ \bibinfo {author} {\bibfnamefont {M.}~\bibnamefont {Mudrich}},\ }\bibfield  {title} {\bibinfo {title} {Ultrafast resonant interatomic coulombic decay induced by quantum fluid dynamics},\ }\href@noop {} {\bibfield  {journal} {\bibinfo  {journal} {Phys. Rev. X}\ }\textbf {\bibinfo {volume} {11}},\ \bibinfo {pages} {021011} (\bibinfo {year} {2021})}\BibitemShut {NoStop}%
\bibitem [{\citenamefont {Asmussen}\ \emph {et~al.}(2021)\citenamefont {Asmussen}, \citenamefont {Michiels}, \citenamefont {Dulitz}, \citenamefont {Ngai}, \citenamefont {Bangert}, \citenamefont {Barranco}, \citenamefont {Binz}, \citenamefont {Bruder}, \citenamefont {Danailov}, \citenamefont {Di~Fraia}, \citenamefont {Eloranta}, \citenamefont {Feifel}, \citenamefont {Giannessi}, \citenamefont {Pi}, \citenamefont {Plekan}, \citenamefont {Prince}, \citenamefont {Squibb}, \citenamefont {Uhl}, \citenamefont {Wituschek}, \citenamefont {Zangrando}, \citenamefont {Callegari}, \citenamefont {Stienkemeier},\ and\ \citenamefont {Mudrich}}]{Asmussen:2021}%
  \BibitemOpen
  \bibfield  {author} {\bibinfo {author} {\bibfnamefont {J.~D.}\ \bibnamefont {Asmussen}}, \bibinfo {author} {\bibfnamefont {R.}~\bibnamefont {Michiels}}, \bibinfo {author} {\bibfnamefont {K.}~\bibnamefont {Dulitz}}, \bibinfo {author} {\bibfnamefont {A.}~\bibnamefont {Ngai}}, \bibinfo {author} {\bibfnamefont {U.}~\bibnamefont {Bangert}}, \bibinfo {author} {\bibfnamefont {M.}~\bibnamefont {Barranco}}, \bibinfo {author} {\bibfnamefont {M.}~\bibnamefont {Binz}}, \bibinfo {author} {\bibfnamefont {L.}~\bibnamefont {Bruder}}, \bibinfo {author} {\bibfnamefont {M.}~\bibnamefont {Danailov}}, \bibinfo {author} {\bibfnamefont {M.}~\bibnamefont {Di~Fraia}}, \bibinfo {author} {\bibfnamefont {J.}~\bibnamefont {Eloranta}}, \bibinfo {author} {\bibfnamefont {R.}~\bibnamefont {Feifel}}, \bibinfo {author} {\bibfnamefont {L.}~\bibnamefont {Giannessi}}, \bibinfo {author} {\bibfnamefont {M.}~\bibnamefont {Pi}}, \bibinfo {author} {\bibfnamefont {O.}~\bibnamefont {Plekan}}, \bibinfo {author} {\bibfnamefont {K.~C.}\ \bibnamefont
  {Prince}}, \bibinfo {author} {\bibfnamefont {R.~J.}\ \bibnamefont {Squibb}}, \bibinfo {author} {\bibfnamefont {D.}~\bibnamefont {Uhl}}, \bibinfo {author} {\bibfnamefont {A.}~\bibnamefont {Wituschek}}, \bibinfo {author} {\bibfnamefont {M.}~\bibnamefont {Zangrando}}, \bibinfo {author} {\bibfnamefont {C.}~\bibnamefont {Callegari}}, \bibinfo {author} {\bibfnamefont {F.}~\bibnamefont {Stienkemeier}},\ and\ \bibinfo {author} {\bibfnamefont {M.}~\bibnamefont {Mudrich}},\ }\bibfield  {title} {\bibinfo {title} {Unravelling the full relaxation dynamics of superexcited helium nanodroplets},\ }\href@noop {} {\bibfield  {journal} {\bibinfo  {journal} {Phys. Chem. Chem. Phys.}\ }\textbf {\bibinfo {volume} {23}},\ \bibinfo {pages} {15138} (\bibinfo {year} {2021})}\BibitemShut {NoStop}%
\bibitem [{\citenamefont {Michiels}\ \emph {et~al.}(2021)\citenamefont {Michiels}, \citenamefont {Abu-samha}, \citenamefont {Madsen}, \citenamefont {Binz}, \citenamefont {Bangert}, \citenamefont {Bruder}, \citenamefont {Duim}, \citenamefont {Wituschek}, \citenamefont {LaForge}, \citenamefont {Squibb}, \citenamefont {Feifel}, \citenamefont {Callegari}, \citenamefont {Di~Fraia}, \citenamefont {Danailov}, \citenamefont {Manfredda}, \citenamefont {Plekan}, \citenamefont {Prince}, \citenamefont {Rebernik}, \citenamefont {Zangrando}, \citenamefont {Stienkemeier},\ and\ \citenamefont {Mudrich}}]{michiels:2021}%
  \BibitemOpen
  \bibfield  {author} {\bibinfo {author} {\bibfnamefont {R.}~\bibnamefont {Michiels}}, \bibinfo {author} {\bibfnamefont {M.}~\bibnamefont {Abu-samha}}, \bibinfo {author} {\bibfnamefont {L.~B.}\ \bibnamefont {Madsen}}, \bibinfo {author} {\bibfnamefont {M.}~\bibnamefont {Binz}}, \bibinfo {author} {\bibfnamefont {U.}~\bibnamefont {Bangert}}, \bibinfo {author} {\bibfnamefont {L.}~\bibnamefont {Bruder}}, \bibinfo {author} {\bibfnamefont {R.}~\bibnamefont {Duim}}, \bibinfo {author} {\bibfnamefont {A.}~\bibnamefont {Wituschek}}, \bibinfo {author} {\bibfnamefont {A.~C.}\ \bibnamefont {LaForge}}, \bibinfo {author} {\bibfnamefont {R.~J.}\ \bibnamefont {Squibb}}, \bibinfo {author} {\bibfnamefont {R.}~\bibnamefont {Feifel}}, \bibinfo {author} {\bibfnamefont {C.}~\bibnamefont {Callegari}}, \bibinfo {author} {\bibfnamefont {M.}~\bibnamefont {Di~Fraia}}, \bibinfo {author} {\bibfnamefont {M.}~\bibnamefont {Danailov}}, \bibinfo {author} {\bibfnamefont {M.}~\bibnamefont {Manfredda}}, \bibinfo {author} {\bibfnamefont
  {O.}~\bibnamefont {Plekan}}, \bibinfo {author} {\bibfnamefont {K.~C.}\ \bibnamefont {Prince}}, \bibinfo {author} {\bibfnamefont {P.}~\bibnamefont {Rebernik}}, \bibinfo {author} {\bibfnamefont {M.}~\bibnamefont {Zangrando}}, \bibinfo {author} {\bibfnamefont {F.}~\bibnamefont {Stienkemeier}},\ and\ \bibinfo {author} {\bibfnamefont {M.}~\bibnamefont {Mudrich}},\ }\bibfield  {title} {\bibinfo {title} {Enhancement of above threshold ionization in resonantly excited helium nanodroplets},\ }\href@noop {} {\bibfield  {journal} {\bibinfo  {journal} {Phys. Rev. Lett.}\ }\textbf {\bibinfo {volume} {127}},\ \bibinfo {pages} {093201} (\bibinfo {year} {2021})}\BibitemShut {NoStop}%
\bibitem [{\citenamefont {LaForge}\ \emph {et~al.}(2022)\citenamefont {LaForge}, \citenamefont {Asmussen}, \citenamefont {Bastian}, \citenamefont {Bonanomi}, \citenamefont {Callegari}, \citenamefont {De}, \citenamefont {Di~Fraia}, \citenamefont {Gorman}, \citenamefont {Hartweg}, \citenamefont {Krishnan} \emph {et~al.}}]{laforge2022relaxation}%
  \BibitemOpen
  \bibfield  {author} {\bibinfo {author} {\bibfnamefont {A.}~\bibnamefont {LaForge}}, \bibinfo {author} {\bibfnamefont {J.~D.}\ \bibnamefont {Asmussen}}, \bibinfo {author} {\bibfnamefont {B.}~\bibnamefont {Bastian}}, \bibinfo {author} {\bibfnamefont {M.}~\bibnamefont {Bonanomi}}, \bibinfo {author} {\bibfnamefont {C.}~\bibnamefont {Callegari}}, \bibinfo {author} {\bibfnamefont {S.}~\bibnamefont {De}}, \bibinfo {author} {\bibfnamefont {M.}~\bibnamefont {Di~Fraia}}, \bibinfo {author} {\bibfnamefont {L.}~\bibnamefont {Gorman}}, \bibinfo {author} {\bibfnamefont {S.}~\bibnamefont {Hartweg}}, \bibinfo {author} {\bibfnamefont {S.}~\bibnamefont {Krishnan}}, \emph {et~al.},\ }\bibfield  {title} {\bibinfo {title} {Relaxation dynamics in excited helium nanodroplets probed with high resolution, time-resolved photoelectron spectroscopy},\ }\href@noop {} {\bibfield  {journal} {\bibinfo  {journal} {Phys. Chem. Chem. Phys.}\ }\textbf {\bibinfo {volume} {24}},\ \bibinfo {pages} {28844} (\bibinfo {year} {2022})}\BibitemShut
  {NoStop}%
\bibitem [{\citenamefont {Buchta}\ \emph {et~al.}(2013{\natexlab{b}})\citenamefont {Buchta}, \citenamefont {Krishnan}, \citenamefont {Brauer}, \citenamefont {Drabbels}, \citenamefont {O'Keeffe}, \citenamefont {Devetta}, \citenamefont {Di~Fraia}, \citenamefont {Callegari}, \citenamefont {Richter}, \citenamefont {Coreno}, \citenamefont {Prince}, \citenamefont {Stienkemeier}, \citenamefont {Moshammer},\ and\ \citenamefont {Mudrich}}]{Buchta:2013}%
  \BibitemOpen
  \bibfield  {author} {\bibinfo {author} {\bibfnamefont {D.}~\bibnamefont {Buchta}}, \bibinfo {author} {\bibfnamefont {S.~R.}\ \bibnamefont {Krishnan}}, \bibinfo {author} {\bibfnamefont {N.~B.}\ \bibnamefont {Brauer}}, \bibinfo {author} {\bibfnamefont {M.}~\bibnamefont {Drabbels}}, \bibinfo {author} {\bibfnamefont {P.}~\bibnamefont {O'Keeffe}}, \bibinfo {author} {\bibfnamefont {M.}~\bibnamefont {Devetta}}, \bibinfo {author} {\bibfnamefont {M.}~\bibnamefont {Di~Fraia}}, \bibinfo {author} {\bibfnamefont {C.}~\bibnamefont {Callegari}}, \bibinfo {author} {\bibfnamefont {R.}~\bibnamefont {Richter}}, \bibinfo {author} {\bibfnamefont {M.}~\bibnamefont {Coreno}}, \bibinfo {author} {\bibfnamefont {K.~C.}\ \bibnamefont {Prince}}, \bibinfo {author} {\bibfnamefont {F.}~\bibnamefont {Stienkemeier}}, \bibinfo {author} {\bibfnamefont {R.}~\bibnamefont {Moshammer}},\ and\ \bibinfo {author} {\bibfnamefont {M.}~\bibnamefont {Mudrich}},\ }\bibfield  {title} {\bibinfo {title} {Charge transfer and penning ionization of dopants in
  or on helium nanodroplets exposed to euv radiation},\ }\href {https://doi.org/10.1021/jp401424w} {\bibfield  {journal} {\bibinfo  {journal} {J. Phys. Chem. A}\ }\textbf {\bibinfo {volume} {117}},\ \bibinfo {pages} {4394} (\bibinfo {year} {2013}{\natexlab{b}})}\BibitemShut {NoStop}%
\bibitem [{\citenamefont {Bouda{\i}ffa}\ \emph {et~al.}(2000)\citenamefont {Bouda{\i}ffa}, \citenamefont {Cloutier}, \citenamefont {Hunting}, \citenamefont {Huels},\ and\ \citenamefont {Sanche}}]{Boudaiffa:2000}%
  \BibitemOpen
  \bibfield  {author} {\bibinfo {author} {\bibfnamefont {B.}~\bibnamefont {Bouda{\i}ffa}}, \bibinfo {author} {\bibfnamefont {P.}~\bibnamefont {Cloutier}}, \bibinfo {author} {\bibfnamefont {D.}~\bibnamefont {Hunting}}, \bibinfo {author} {\bibfnamefont {M.~A.}\ \bibnamefont {Huels}},\ and\ \bibinfo {author} {\bibfnamefont {L.}~\bibnamefont {Sanche}},\ }\bibfield  {title} {\bibinfo {title} {Resonant formation of dna strand breaks by low-energy (3 to 20 ev) electrons},\ }\href@noop {} {\bibfield  {journal} {\bibinfo  {journal} {Science}\ }\textbf {\bibinfo {volume} {287}},\ \bibinfo {pages} {1658} (\bibinfo {year} {2000})}\BibitemShut {NoStop}%
\bibitem [{\citenamefont {Alizadeh}\ \emph {et~al.}(2015)\citenamefont {Alizadeh}, \citenamefont {Orlando},\ and\ \citenamefont {Sanche}}]{Alizadeh:2015}%
  \BibitemOpen
  \bibfield  {author} {\bibinfo {author} {\bibfnamefont {E.}~\bibnamefont {Alizadeh}}, \bibinfo {author} {\bibfnamefont {T.~M.}\ \bibnamefont {Orlando}},\ and\ \bibinfo {author} {\bibfnamefont {L.}~\bibnamefont {Sanche}},\ }\bibfield  {title} {\bibinfo {title} {Biomolecular damage induced by ionizing radiation: The direct and indirect effects of low-energy electrons on dna},\ }\href@noop {} {\bibfield  {journal} {\bibinfo  {journal} {Annu. Rev. Phys. Chem.}\ }\textbf {\bibinfo {volume} {66}},\ \bibinfo {pages} {379} (\bibinfo {year} {2015})}\BibitemShut {NoStop}%
\bibitem [{\citenamefont {Pimblott}\ and\ \citenamefont {LaVerne}(2007)}]{pimblott2007}%
  \BibitemOpen
  \bibfield  {author} {\bibinfo {author} {\bibfnamefont {S.~M.}\ \bibnamefont {Pimblott}}\ and\ \bibinfo {author} {\bibfnamefont {J.~A.}\ \bibnamefont {LaVerne}},\ }\bibfield  {title} {\bibinfo {title} {Production of low-energy electrons by ionizing radiation},\ }\href@noop {} {\bibfield  {journal} {\bibinfo  {journal} {Radiat. Phys. Chem.}\ }\textbf {\bibinfo {volume} {76}},\ \bibinfo {pages} {1244} (\bibinfo {year} {2007})}\BibitemShut {NoStop}%
\bibitem [{\citenamefont {Barth}\ \emph {et~al.}(2005)\citenamefont {Barth}, \citenamefont {Joshi}, \citenamefont {Marburger}, \citenamefont {Ulrich}, \citenamefont {Lindblad}, \citenamefont {{\"O}hrwall}, \citenamefont {Bj{\"o}rneholm},\ and\ \citenamefont {Hergenhahn}}]{barth2005}%
  \BibitemOpen
  \bibfield  {author} {\bibinfo {author} {\bibfnamefont {S.}~\bibnamefont {Barth}}, \bibinfo {author} {\bibfnamefont {S.}~\bibnamefont {Joshi}}, \bibinfo {author} {\bibfnamefont {S.}~\bibnamefont {Marburger}}, \bibinfo {author} {\bibfnamefont {V.}~\bibnamefont {Ulrich}}, \bibinfo {author} {\bibfnamefont {A.}~\bibnamefont {Lindblad}}, \bibinfo {author} {\bibfnamefont {G.}~\bibnamefont {{\"O}hrwall}}, \bibinfo {author} {\bibfnamefont {O.}~\bibnamefont {Bj{\"o}rneholm}},\ and\ \bibinfo {author} {\bibfnamefont {U.}~\bibnamefont {Hergenhahn}},\ }\bibfield  {title} {\bibinfo {title} {Observation of resonant interatomic coulombic decay in ne clusters},\ }\href@noop {} {\bibfield  {journal} {\bibinfo  {journal} {J. Chem. Phys.}\ }\textbf {\bibinfo {volume} {122}} (\bibinfo {year} {2005})}\BibitemShut {NoStop}%
\bibitem [{\citenamefont {Mucke}\ \emph {et~al.}(2015)\citenamefont {Mucke}, \citenamefont {Arion}, \citenamefont {F{\"o}rstel}, \citenamefont {Lischke},\ and\ \citenamefont {Hergenhahn}}]{mucke2015}%
  \BibitemOpen
  \bibfield  {author} {\bibinfo {author} {\bibfnamefont {M.}~\bibnamefont {Mucke}}, \bibinfo {author} {\bibfnamefont {T.}~\bibnamefont {Arion}}, \bibinfo {author} {\bibfnamefont {M.}~\bibnamefont {F{\"o}rstel}}, \bibinfo {author} {\bibfnamefont {T.}~\bibnamefont {Lischke}},\ and\ \bibinfo {author} {\bibfnamefont {U.}~\bibnamefont {Hergenhahn}},\ }\bibfield  {title} {\bibinfo {title} {Competition of inelastic electron scattering and interatomic coulombic decay in ne clusters},\ }\href@noop {} {\bibfield  {journal} {\bibinfo  {journal} {J. Electron Spectros. Relat. Phenomena}\ }\textbf {\bibinfo {volume} {200}},\ \bibinfo {pages} {232} (\bibinfo {year} {2015})}\BibitemShut {NoStop}%
\bibitem [{\citenamefont {Iablonskyi}\ \emph {et~al.}(2016)\citenamefont {Iablonskyi}, \citenamefont {Nagaya}, \citenamefont {Fukuzawa}, \citenamefont {Motomura}, \citenamefont {Kumagai}, \citenamefont {Mondal}, \citenamefont {Tachibana}, \citenamefont {Takanashi}, \citenamefont {Nishiyama}, \citenamefont {Matsunami}, \citenamefont {Johnsson}, \citenamefont {Piseri}, \citenamefont {Sansone}, \citenamefont {Dubrouil}, \citenamefont {Reduzzi}, \citenamefont {Carpeggiani}, \citenamefont {Vozzi}, \citenamefont {Devetta}, \citenamefont {Negro}, \citenamefont {Calegari}, \citenamefont {Trabattoni}, \citenamefont {Castrovilli}, \citenamefont {Faccial\`a}, \citenamefont {Ovcharenko}, \citenamefont {M\"oller}, \citenamefont {Mudrich}, \citenamefont {Stienkemeier}, \citenamefont {Coreno}, \citenamefont {Alagia}, \citenamefont {Sch\"utte}, \citenamefont {Berrah}, \citenamefont {Kuleff}, \citenamefont {Jabbari}, \citenamefont {Callegari}, \citenamefont {Plekan}, \citenamefont {Finetti}, \citenamefont {Spezzani},
  \citenamefont {Ferrari}, \citenamefont {Allaria}, \citenamefont {Penco}, \citenamefont {Serpico}, \citenamefont {De~Ninno}, \citenamefont {Nikolov}, \citenamefont {Diviacco}, \citenamefont {Di~Mitri}, \citenamefont {Giannessi}, \citenamefont {Prince},\ and\ \citenamefont {Ueda}}]{Iablonskyi:2016}%
  \BibitemOpen
  \bibfield  {author} {\bibinfo {author} {\bibfnamefont {D.}~\bibnamefont {Iablonskyi}}, \bibinfo {author} {\bibfnamefont {K.}~\bibnamefont {Nagaya}}, \bibinfo {author} {\bibfnamefont {H.}~\bibnamefont {Fukuzawa}}, \bibinfo {author} {\bibfnamefont {K.}~\bibnamefont {Motomura}}, \bibinfo {author} {\bibfnamefont {Y.}~\bibnamefont {Kumagai}}, \bibinfo {author} {\bibfnamefont {S.}~\bibnamefont {Mondal}}, \bibinfo {author} {\bibfnamefont {T.}~\bibnamefont {Tachibana}}, \bibinfo {author} {\bibfnamefont {T.}~\bibnamefont {Takanashi}}, \bibinfo {author} {\bibfnamefont {T.}~\bibnamefont {Nishiyama}}, \bibinfo {author} {\bibfnamefont {K.}~\bibnamefont {Matsunami}}, \bibinfo {author} {\bibfnamefont {P.}~\bibnamefont {Johnsson}}, \bibinfo {author} {\bibfnamefont {P.}~\bibnamefont {Piseri}}, \bibinfo {author} {\bibfnamefont {G.}~\bibnamefont {Sansone}}, \bibinfo {author} {\bibfnamefont {A.}~\bibnamefont {Dubrouil}}, \bibinfo {author} {\bibfnamefont {M.}~\bibnamefont {Reduzzi}}, \bibinfo {author} {\bibfnamefont
  {P.}~\bibnamefont {Carpeggiani}}, \bibinfo {author} {\bibfnamefont {C.}~\bibnamefont {Vozzi}}, \bibinfo {author} {\bibfnamefont {M.}~\bibnamefont {Devetta}}, \bibinfo {author} {\bibfnamefont {M.}~\bibnamefont {Negro}}, \bibinfo {author} {\bibfnamefont {F.}~\bibnamefont {Calegari}}, \bibinfo {author} {\bibfnamefont {A.}~\bibnamefont {Trabattoni}}, \bibinfo {author} {\bibfnamefont {M.~C.}\ \bibnamefont {Castrovilli}}, \bibinfo {author} {\bibfnamefont {D.}~\bibnamefont {Faccial\`a}}, \bibinfo {author} {\bibfnamefont {Y.}~\bibnamefont {Ovcharenko}}, \bibinfo {author} {\bibfnamefont {T.}~\bibnamefont {M\"oller}}, \bibinfo {author} {\bibfnamefont {M.}~\bibnamefont {Mudrich}}, \bibinfo {author} {\bibfnamefont {F.}~\bibnamefont {Stienkemeier}}, \bibinfo {author} {\bibfnamefont {M.}~\bibnamefont {Coreno}}, \bibinfo {author} {\bibfnamefont {M.}~\bibnamefont {Alagia}}, \bibinfo {author} {\bibfnamefont {B.}~\bibnamefont {Sch\"utte}}, \bibinfo {author} {\bibfnamefont {N.}~\bibnamefont {Berrah}}, \bibinfo {author}
  {\bibfnamefont {A.~I.}\ \bibnamefont {Kuleff}}, \bibinfo {author} {\bibfnamefont {G.}~\bibnamefont {Jabbari}}, \bibinfo {author} {\bibfnamefont {C.}~\bibnamefont {Callegari}}, \bibinfo {author} {\bibfnamefont {O.}~\bibnamefont {Plekan}}, \bibinfo {author} {\bibfnamefont {P.}~\bibnamefont {Finetti}}, \bibinfo {author} {\bibfnamefont {C.}~\bibnamefont {Spezzani}}, \bibinfo {author} {\bibfnamefont {E.}~\bibnamefont {Ferrari}}, \bibinfo {author} {\bibfnamefont {E.}~\bibnamefont {Allaria}}, \bibinfo {author} {\bibfnamefont {G.}~\bibnamefont {Penco}}, \bibinfo {author} {\bibfnamefont {C.}~\bibnamefont {Serpico}}, \bibinfo {author} {\bibfnamefont {G.}~\bibnamefont {De~Ninno}}, \bibinfo {author} {\bibfnamefont {I.}~\bibnamefont {Nikolov}}, \bibinfo {author} {\bibfnamefont {B.}~\bibnamefont {Diviacco}}, \bibinfo {author} {\bibfnamefont {S.}~\bibnamefont {Di~Mitri}}, \bibinfo {author} {\bibfnamefont {L.}~\bibnamefont {Giannessi}}, \bibinfo {author} {\bibfnamefont {K.~C.}\ \bibnamefont {Prince}},\ and\ \bibinfo
  {author} {\bibfnamefont {K.}~\bibnamefont {Ueda}},\ }\bibfield  {title} {\bibinfo {title} {Slow interatomic coulombic decay of multiply excited neon clusters},\ }\href@noop {} {\bibfield  {journal} {\bibinfo  {journal} {Phys. Rev. Lett.}\ }\textbf {\bibinfo {volume} {117}},\ \bibinfo {pages} {276806} (\bibinfo {year} {2016})}\BibitemShut {NoStop}%
\bibitem [{\citenamefont {Ltaief}\ \emph {et~al.}(2018)\citenamefont {Ltaief}, \citenamefont {Hans}, \citenamefont {Schmidt}, \citenamefont {Holzapfel}, \citenamefont {Wiegandt}, \citenamefont {Reiss}, \citenamefont {K{\"u}stner-Wetekam}, \citenamefont {Jahnke}, \citenamefont {D{\"o}rner}, \citenamefont {Knie},\ and\ \citenamefont {Ehresmann}}]{ltaief:2018}%
  \BibitemOpen
  \bibfield  {author} {\bibinfo {author} {\bibfnamefont {L.~B.}\ \bibnamefont {Ltaief}}, \bibinfo {author} {\bibfnamefont {A.}~\bibnamefont {Hans}}, \bibinfo {author} {\bibfnamefont {P.}~\bibnamefont {Schmidt}}, \bibinfo {author} {\bibfnamefont {X.}~\bibnamefont {Holzapfel}}, \bibinfo {author} {\bibfnamefont {F.}~\bibnamefont {Wiegandt}}, \bibinfo {author} {\bibfnamefont {P.}~\bibnamefont {Reiss}}, \bibinfo {author} {\bibfnamefont {C.}~\bibnamefont {K{\"u}stner-Wetekam}}, \bibinfo {author} {\bibfnamefont {T.}~\bibnamefont {Jahnke}}, \bibinfo {author} {\bibfnamefont {R.}~\bibnamefont {D{\"o}rner}}, \bibinfo {author} {\bibfnamefont {A.}~\bibnamefont {Knie}},\ and\ \bibinfo {author} {\bibfnamefont {A.}~\bibnamefont {Ehresmann}},\ }\bibfield  {title} {\bibinfo {title} {Vuv photon emission from ne clusters of varying sizes following photon and photoelectron excitations},\ }\href@noop {} {\bibfield  {journal} {\bibinfo  {journal} {J. Phys. B: At., Mol. Opt. Phys.}\ }\textbf {\bibinfo {volume} {51}},\ \bibinfo
  {pages} {065002} (\bibinfo {year} {2018})}\BibitemShut {NoStop}%
\bibitem [{\citenamefont {Malerz}\ \emph {et~al.}(2021)\citenamefont {Malerz}, \citenamefont {Trinter}, \citenamefont {Hergenhahn}, \citenamefont {Ghrist}, \citenamefont {Ali}, \citenamefont {Nicolas}, \citenamefont {Saak}, \citenamefont {Richter}, \citenamefont {Hartweg}, \citenamefont {Nahon} \emph {et~al.}}]{malerz2021low}%
  \BibitemOpen
  \bibfield  {author} {\bibinfo {author} {\bibfnamefont {S.}~\bibnamefont {Malerz}}, \bibinfo {author} {\bibfnamefont {F.}~\bibnamefont {Trinter}}, \bibinfo {author} {\bibfnamefont {U.}~\bibnamefont {Hergenhahn}}, \bibinfo {author} {\bibfnamefont {A.}~\bibnamefont {Ghrist}}, \bibinfo {author} {\bibfnamefont {H.}~\bibnamefont {Ali}}, \bibinfo {author} {\bibfnamefont {C.}~\bibnamefont {Nicolas}}, \bibinfo {author} {\bibfnamefont {C.-M.}\ \bibnamefont {Saak}}, \bibinfo {author} {\bibfnamefont {C.}~\bibnamefont {Richter}}, \bibinfo {author} {\bibfnamefont {S.}~\bibnamefont {Hartweg}}, \bibinfo {author} {\bibfnamefont {L.}~\bibnamefont {Nahon}}, \emph {et~al.},\ }\bibfield  {title} {\bibinfo {title} {Low-energy constraints on photoelectron spectra measured from liquid water and aqueous solutions},\ }\href@noop {} {\bibfield  {journal} {\bibinfo  {journal} {Phys. Chem. Chem. Phys.}\ }\textbf {\bibinfo {volume} {23}},\ \bibinfo {pages} {8246} (\bibinfo {year} {2021})}\BibitemShut {NoStop}%
\bibitem [{\citenamefont {Toennies}\ \emph {et~al.}(2001)\citenamefont {Toennies}, \citenamefont {Vilesov},\ and\ \citenamefont {Whaley}}]{Toennies:2001}%
  \BibitemOpen
  \bibfield  {author} {\bibinfo {author} {\bibfnamefont {J.~P.}\ \bibnamefont {Toennies}}, \bibinfo {author} {\bibfnamefont {A.~F.}\ \bibnamefont {Vilesov}},\ and\ \bibinfo {author} {\bibfnamefont {K.~B.}\ \bibnamefont {Whaley}},\ }\bibfield  {title} {\bibinfo {title} {Superfluid helium droplets: An ultracold nanolaboratory},\ }\href@noop {} {\bibfield  {journal} {\bibinfo  {journal} {Physics Today}\ }\textbf {\bibinfo {volume} {45}},\ \bibinfo {pages} {31} (\bibinfo {year} {2001})}\BibitemShut {NoStop}%
\bibitem [{\citenamefont {Bastian}\ \emph {et~al.}(2022)\citenamefont {Bastian}, \citenamefont {Asmussen}, \citenamefont {Ben~Ltaief}, \citenamefont {Czasch}, \citenamefont {Jones}, \citenamefont {Hoffmann}, \citenamefont {Pedersen},\ and\ \citenamefont {Mudrich}}]{bastian2022}%
  \BibitemOpen
  \bibfield  {author} {\bibinfo {author} {\bibfnamefont {B.}~\bibnamefont {Bastian}}, \bibinfo {author} {\bibfnamefont {J.~D.}\ \bibnamefont {Asmussen}}, \bibinfo {author} {\bibfnamefont {L.}~\bibnamefont {Ben~Ltaief}}, \bibinfo {author} {\bibfnamefont {A.}~\bibnamefont {Czasch}}, \bibinfo {author} {\bibfnamefont {N.~C.}\ \bibnamefont {Jones}}, \bibinfo {author} {\bibfnamefont {S.~V.}\ \bibnamefont {Hoffmann}}, \bibinfo {author} {\bibfnamefont {H.~B.}\ \bibnamefont {Pedersen}},\ and\ \bibinfo {author} {\bibfnamefont {M.}~\bibnamefont {Mudrich}},\ }\bibfield  {title} {\bibinfo {title} {A new endstation for extreme-ultraviolet spectroscopy of free clusters and nanodroplets},\ }\href@noop {} {\bibfield  {journal} {\bibinfo  {journal} {Rev. Sci. Instrum.}\ }\textbf {\bibinfo {volume} {93}} (\bibinfo {year} {2022})}\BibitemShut {NoStop}%
\bibitem [{\citenamefont {Kuma}\ \emph {et~al.}(2007)\citenamefont {Kuma}, \citenamefont {Goto}, \citenamefont {Slipchenko}, \citenamefont {Vilesov}, \citenamefont {Khramov},\ and\ \citenamefont {Momose}}]{kuma2007laser}%
  \BibitemOpen
  \bibfield  {author} {\bibinfo {author} {\bibfnamefont {S.}~\bibnamefont {Kuma}}, \bibinfo {author} {\bibfnamefont {H.}~\bibnamefont {Goto}}, \bibinfo {author} {\bibfnamefont {M.~N.}\ \bibnamefont {Slipchenko}}, \bibinfo {author} {\bibfnamefont {A.~F.}\ \bibnamefont {Vilesov}}, \bibinfo {author} {\bibfnamefont {A.}~\bibnamefont {Khramov}},\ and\ \bibinfo {author} {\bibfnamefont {T.}~\bibnamefont {Momose}},\ }\bibfield  {title} {\bibinfo {title} {Laser induced fluorescence of mg-phthalocyanine in he droplets: Evidence for fluxionality of large {H}$_2$ clusters at 0.38 {K}},\ }\href@noop {} {\bibfield  {journal} {\bibinfo  {journal} {J. Chem. Phys.}\ }\textbf {\bibinfo {volume} {127}} (\bibinfo {year} {2007})}\BibitemShut {NoStop}%
\bibitem [{\citenamefont {B{\"u}nermann}\ and\ \citenamefont {Stienkemeier}(2011)}]{Buenermann:2011}%
  \BibitemOpen
  \bibfield  {author} {\bibinfo {author} {\bibfnamefont {O.}~\bibnamefont {B{\"u}nermann}}\ and\ \bibinfo {author} {\bibfnamefont {F.}~\bibnamefont {Stienkemeier}},\ }\bibfield  {title} {\bibinfo {title} {Modeling the formation of alkali clusters attached to helium nanodroplets and the abundance of high-spin states},\ }\href@noop {} {\bibfield  {journal} {\bibinfo  {journal} {Eur. Phys. J. D}\ }\textbf {\bibinfo {volume} {61}},\ \bibinfo {pages} {645} (\bibinfo {year} {2011})}\BibitemShut {NoStop}%
\bibitem [{\citenamefont {Dick}(2014)}]{Dick:2014}%
  \BibitemOpen
  \bibfield  {author} {\bibinfo {author} {\bibfnamefont {B.}~\bibnamefont {Dick}},\ }\bibfield  {title} {\bibinfo {title} {Inverting ion images without abel inversion: maximum entropy reconstruction of velocity maps},\ }\href@noop {} {\bibfield  {journal} {\bibinfo  {journal} {Phys. Chem. Chem. Phys.}\ }\textbf {\bibinfo {volume} {16}},\ \bibinfo {pages} {570} (\bibinfo {year} {2014})}\BibitemShut {NoStop}%
\bibitem [{\citenamefont {Joppien}\ \emph {et~al.}(1993)\citenamefont {Joppien}, \citenamefont {Karnbach},\ and\ \citenamefont {M\"oller}}]{Joppien:1993}%
  \BibitemOpen
  \bibfield  {author} {\bibinfo {author} {\bibfnamefont {M.}~\bibnamefont {Joppien}}, \bibinfo {author} {\bibfnamefont {R.}~\bibnamefont {Karnbach}},\ and\ \bibinfo {author} {\bibfnamefont {T.}~\bibnamefont {M\"oller}},\ }\bibfield  {title} {\bibinfo {title} {Electronic excitations in liquid helium: The evolution from small clusters to large droplets},\ }\href@noop {} {\bibfield  {journal} {\bibinfo  {journal} {Phys. Rev. Lett.}\ }\textbf {\bibinfo {volume} {71}},\ \bibinfo {pages} {2654} (\bibinfo {year} {1993})}\BibitemShut {NoStop}%
\bibitem [{\citenamefont {Scheidemann}\ \emph {et~al.}(1997)\citenamefont {Scheidemann}, \citenamefont {Kresin},\ and\ \citenamefont {Hess}}]{Scheidemann:1997}%
  \BibitemOpen
  \bibfield  {author} {\bibinfo {author} {\bibfnamefont {A.~A.}\ \bibnamefont {Scheidemann}}, \bibinfo {author} {\bibfnamefont {V.~V.}\ \bibnamefont {Kresin}},\ and\ \bibinfo {author} {\bibfnamefont {H.}~\bibnamefont {Hess}},\ }\bibfield  {title} {\bibinfo {title} {Capture of lithium by $^4${H}e clusters: Surface adsorption, penning ionization, and formation of {H}e{L}i$^+$},\ }\href@noop {} {\bibfield  {journal} {\bibinfo  {journal} {J. Chem. Phys.}\ }\textbf {\bibinfo {volume} {107}},\ \bibinfo {pages} {2839} (\bibinfo {year} {1997})}\BibitemShut {NoStop}%
\bibitem [{\citenamefont {McKinsey}\ \emph {et~al.}(2003)\citenamefont {McKinsey}, \citenamefont {Brome}, \citenamefont {Dzhosyuk}, \citenamefont {Golub}, \citenamefont {Habicht}, \citenamefont {Huffman}, \citenamefont {Korobkina}, \citenamefont {Lamoreaux}, \citenamefont {Mattoni}, \citenamefont {Thompson}, \citenamefont {Yang},\ and\ \citenamefont {Doyle}}]{McKinsey:2003}%
  \BibitemOpen
  \bibfield  {author} {\bibinfo {author} {\bibfnamefont {D.~N.}\ \bibnamefont {McKinsey}}, \bibinfo {author} {\bibfnamefont {C.~R.}\ \bibnamefont {Brome}}, \bibinfo {author} {\bibfnamefont {S.~N.}\ \bibnamefont {Dzhosyuk}}, \bibinfo {author} {\bibfnamefont {R.}~\bibnamefont {Golub}}, \bibinfo {author} {\bibfnamefont {K.}~\bibnamefont {Habicht}}, \bibinfo {author} {\bibfnamefont {P.~R.}\ \bibnamefont {Huffman}}, \bibinfo {author} {\bibfnamefont {E.}~\bibnamefont {Korobkina}}, \bibinfo {author} {\bibfnamefont {S.~K.}\ \bibnamefont {Lamoreaux}}, \bibinfo {author} {\bibfnamefont {C.~E.~H.}\ \bibnamefont {Mattoni}}, \bibinfo {author} {\bibfnamefont {A.~K.}\ \bibnamefont {Thompson}}, \bibinfo {author} {\bibfnamefont {L.}~\bibnamefont {Yang}},\ and\ \bibinfo {author} {\bibfnamefont {J.~M.}\ \bibnamefont {Doyle}},\ }\bibfield  {title} {\bibinfo {title} {Time dependence of liquid-helium fluorescence},\ }\href@noop {} {\bibfield  {journal} {\bibinfo  {journal} {Phys. Rev. A}\ }\textbf {\bibinfo {volume} {67}},\
  \bibinfo {pages} {062716} (\bibinfo {year} {2003})}\BibitemShut {NoStop}%
\bibitem [{\citenamefont {Haberland}\ \emph {et~al.}(1995)\citenamefont {Haberland}, \citenamefont {Issendorff}, \citenamefont {Fr{\"o}chtenicht},\ and\ \citenamefont {Toennies}}]{haberland1995absorption}%
  \BibitemOpen
  \bibfield  {author} {\bibinfo {author} {\bibfnamefont {H.}~\bibnamefont {Haberland}}, \bibinfo {author} {\bibfnamefont {B.~v.}\ \bibnamefont {Issendorff}}, \bibinfo {author} {\bibfnamefont {R.}~\bibnamefont {Fr{\"o}chtenicht}},\ and\ \bibinfo {author} {\bibfnamefont {J.}~\bibnamefont {Toennies}},\ }\bibfield  {title} {\bibinfo {title} {Absorption spectroscopy and photodissociation dynamics of small helium cluster ions},\ }\href@noop {} {\bibfield  {journal} {\bibinfo  {journal} {J. Chem. Phys.}\ }\textbf {\bibinfo {volume} {102}},\ \bibinfo {pages} {8773} (\bibinfo {year} {1995})}\BibitemShut {NoStop}%
\bibitem [{\citenamefont {Samson}\ and\ \citenamefont {Stolte}(2002)}]{Samson:2002}%
  \BibitemOpen
  \bibfield  {author} {\bibinfo {author} {\bibfnamefont {J.}~\bibnamefont {Samson}}\ and\ \bibinfo {author} {\bibfnamefont {W.~C.}\ \bibnamefont {Stolte}},\ }\bibfield  {title} {\bibinfo {title} {Precision measurements of the total photoionization cross-sections of {H}e, {N}e, {A}r, {K}r, and {X}e},\ }\href@noop {} {\bibfield  {journal} {\bibinfo  {journal} {J. Electron. Spectros. Relat. Phenomena}\ }\textbf {\bibinfo {volume} {123}},\ \bibinfo {pages} {265} (\bibinfo {year} {2002})}\BibitemShut {NoStop}%
\bibitem [{\citenamefont {Asmussen}\ \emph {et~al.}(2023{\natexlab{c}})\citenamefont {Asmussen}, \citenamefont {Sishodia}, \citenamefont {Bastian}, \citenamefont {Abid}, \citenamefont {Ltaief}, \citenamefont {Pedersen}, \citenamefont {De}, \citenamefont {Medina}, \citenamefont {Pal}, \citenamefont {Richter} \emph {et~al.}}]{asmussen2023electron}%
  \BibitemOpen
  \bibfield  {author} {\bibinfo {author} {\bibfnamefont {J.~D.}\ \bibnamefont {Asmussen}}, \bibinfo {author} {\bibfnamefont {K.}~\bibnamefont {Sishodia}}, \bibinfo {author} {\bibfnamefont {B.}~\bibnamefont {Bastian}}, \bibinfo {author} {\bibfnamefont {A.~R.}\ \bibnamefont {Abid}}, \bibinfo {author} {\bibfnamefont {L.~B.}\ \bibnamefont {Ltaief}}, \bibinfo {author} {\bibfnamefont {H.~B.}\ \bibnamefont {Pedersen}}, \bibinfo {author} {\bibfnamefont {S.}~\bibnamefont {De}}, \bibinfo {author} {\bibfnamefont {C.}~\bibnamefont {Medina}}, \bibinfo {author} {\bibfnamefont {N.}~\bibnamefont {Pal}}, \bibinfo {author} {\bibfnamefont {R.}~\bibnamefont {Richter}}, \emph {et~al.},\ }\bibfield  {title} {\bibinfo {title} {Electron energy loss and angular asymmetry induced by elastic scattering in superfluid helium nanodroplets},\ }\href@noop {} {\bibfield  {journal} {\bibinfo  {journal} {Nanoscale}\ }\textbf {\bibinfo {volume} {15}},\ \bibinfo {pages} {14025} (\bibinfo {year} {2023}{\natexlab{c}})}\BibitemShut {NoStop}%
\bibitem [{\citenamefont {Shcherbinin}\ \emph {et~al.}(2019)\citenamefont {Shcherbinin}, \citenamefont {Westergaard}, \citenamefont {Hanif}, \citenamefont {Krishnan}, \citenamefont {LaForge}, \citenamefont {Richter}, \citenamefont {Pfeifer},\ and\ \citenamefont {Mudrich}}]{Shcherbinin:2019}%
  \BibitemOpen
  \bibfield  {author} {\bibinfo {author} {\bibfnamefont {M.}~\bibnamefont {Shcherbinin}}, \bibinfo {author} {\bibfnamefont {F.~V.}\ \bibnamefont {Westergaard}}, \bibinfo {author} {\bibfnamefont {M.}~\bibnamefont {Hanif}}, \bibinfo {author} {\bibfnamefont {S.}~\bibnamefont {Krishnan}}, \bibinfo {author} {\bibfnamefont {A.}~\bibnamefont {LaForge}}, \bibinfo {author} {\bibfnamefont {R.}~\bibnamefont {Richter}}, \bibinfo {author} {\bibfnamefont {T.}~\bibnamefont {Pfeifer}},\ and\ \bibinfo {author} {\bibfnamefont {M.}~\bibnamefont {Mudrich}},\ }\bibfield  {title} {\bibinfo {title} {Inelastic scattering of photoelectrons from he nanodroplets},\ }\href@noop {} {\bibfield  {journal} {\bibinfo  {journal} {J. Chem. Phys.}\ }\textbf {\bibinfo {volume} {150}},\ \bibinfo {pages} {044304} (\bibinfo {year} {2019})}\BibitemShut {NoStop}%
\bibitem [{\citenamefont {Buchenau}\ \emph {et~al.}(1991)\citenamefont {Buchenau}, \citenamefont {Toennies},\ and\ \citenamefont {Northby}}]{Buchenau:1991}%
  \BibitemOpen
  \bibfield  {author} {\bibinfo {author} {\bibfnamefont {H.}~\bibnamefont {Buchenau}}, \bibinfo {author} {\bibfnamefont {J.~P.}\ \bibnamefont {Toennies}},\ and\ \bibinfo {author} {\bibfnamefont {J.~A.}\ \bibnamefont {Northby}},\ }\bibfield  {title} {\bibinfo {title} {Excitation and ionization of $^4$he clusters by electrons},\ }\href@noop {} {\bibfield  {journal} {\bibinfo  {journal} {J. Chem. Phys.}\ }\textbf {\bibinfo {volume} {95}},\ \bibinfo {pages} {8134} (\bibinfo {year} {1991})}\BibitemShut {NoStop}%
\bibitem [{\citenamefont {Fiedler}\ and\ \citenamefont {Eloranta}(2014)}]{Fiedler:2014}%
  \BibitemOpen
  \bibfield  {author} {\bibinfo {author} {\bibfnamefont {S.~L.}\ \bibnamefont {Fiedler}}\ and\ \bibinfo {author} {\bibfnamefont {J.}~\bibnamefont {Eloranta}},\ }\bibfield  {title} {\bibinfo {title} {Interaction of helium rydberg state atoms with superfluid helium},\ }\href@noop {} {\bibfield  {journal} {\bibinfo  {journal} {J. Low Temp. Phys.}\ }\textbf {\bibinfo {volume} {174}},\ \bibinfo {pages} {269} (\bibinfo {year} {2014})}\BibitemShut {NoStop}%
\bibitem [{\citenamefont {Nijjar}\ \emph {et~al.}(2018)\citenamefont {Nijjar}, \citenamefont {Krylov}, \citenamefont {Prezhdo}, \citenamefont {Vilesov},\ and\ \citenamefont {Wittig}}]{nijjar2018conversion}%
  \BibitemOpen
  \bibfield  {author} {\bibinfo {author} {\bibfnamefont {P.}~\bibnamefont {Nijjar}}, \bibinfo {author} {\bibfnamefont {A.}~\bibnamefont {Krylov}}, \bibinfo {author} {\bibfnamefont {O.}~\bibnamefont {Prezhdo}}, \bibinfo {author} {\bibfnamefont {A.}~\bibnamefont {Vilesov}},\ and\ \bibinfo {author} {\bibfnamefont {C.}~\bibnamefont {Wittig}},\ }\bibfield  {title} {\bibinfo {title} {Conversion of {H}e (2$^3${S}) to {H}e$_2$ ({a}$^3\sigma_u^+$) in liquid helium},\ }\href@noop {} {\bibfield  {journal} {\bibinfo  {journal} {J. Phys. Chem. Lett.}\ }\textbf {\bibinfo {volume} {9}},\ \bibinfo {pages} {6017} (\bibinfo {year} {2018})}\BibitemShut {NoStop}%
\bibitem [{\citenamefont {Asmussen}\ \emph {et~al.}(2022)\citenamefont {Asmussen}, \citenamefont {Michiels}, \citenamefont {Bangert}, \citenamefont {Sisourat}, \citenamefont {Binz}, \citenamefont {Bruder}, \citenamefont {Danailov}, \citenamefont {Di~Fraia}, \citenamefont {Feifel}, \citenamefont {Giannessi} \emph {et~al.}}]{Asmussen:2022}%
  \BibitemOpen
  \bibfield  {author} {\bibinfo {author} {\bibfnamefont {J.~D.}\ \bibnamefont {Asmussen}}, \bibinfo {author} {\bibfnamefont {R.}~\bibnamefont {Michiels}}, \bibinfo {author} {\bibfnamefont {U.}~\bibnamefont {Bangert}}, \bibinfo {author} {\bibfnamefont {N.}~\bibnamefont {Sisourat}}, \bibinfo {author} {\bibfnamefont {M.}~\bibnamefont {Binz}}, \bibinfo {author} {\bibfnamefont {L.}~\bibnamefont {Bruder}}, \bibinfo {author} {\bibfnamefont {M.}~\bibnamefont {Danailov}}, \bibinfo {author} {\bibfnamefont {M.}~\bibnamefont {Di~Fraia}}, \bibinfo {author} {\bibfnamefont {R.}~\bibnamefont {Feifel}}, \bibinfo {author} {\bibfnamefont {L.}~\bibnamefont {Giannessi}}, \emph {et~al.},\ }\bibfield  {title} {\bibinfo {title} {Time-resolved ultrafast interatomic coulombic decay in superexcited sodium-doped helium nanodroplets},\ }\href@noop {} {\bibfield  {journal} {\bibinfo  {journal} {J. Phys. Chem. Lett.}\ }\textbf {\bibinfo {volume} {13}},\ \bibinfo {pages} {4470} (\bibinfo {year} {2022})}\BibitemShut {NoStop}%
\bibitem [{Note1()}]{Note1}%
  \BibitemOpen
  \bibinfo {note} {This estimate of the time constant is based on the assumption that two He$^*$'s emerging to the surface of He nanodroplets undergo a roaming motion before colliding and decaying by ICD. The roaming velocity is taken as the critical Landau velocity $\approx 60$~m/s~\cite {Brauer:2013} and the mean roaming distance is $\sim R$.}\BibitemShut {Stop}%
\bibitem [{\citenamefont {Benderskii}\ \emph {et~al.}(1999)\citenamefont {Benderskii}, \citenamefont {Zadoyan}, \citenamefont {Schwentner},\ and\ \citenamefont {Apkarian}}]{Benderskii:1999}%
  \BibitemOpen
  \bibfield  {author} {\bibinfo {author} {\bibfnamefont {A.~V.}\ \bibnamefont {Benderskii}}, \bibinfo {author} {\bibfnamefont {R.}~\bibnamefont {Zadoyan}}, \bibinfo {author} {\bibfnamefont {N.}~\bibnamefont {Schwentner}},\ and\ \bibinfo {author} {\bibfnamefont {V.~A.}\ \bibnamefont {Apkarian}},\ }\bibfield  {title} {\bibinfo {title} {Photodynamics in superfluid helium: Femtosecond laser-induced ionization, charge recombination, and preparation of molecular {R}ydberg states},\ }\href@noop {} {\bibfield  {journal} {\bibinfo  {journal} {J. Chem. Phys.}\ }\textbf {\bibinfo {volume} {110}},\ \bibinfo {pages} {1542} (\bibinfo {year} {1999})}\BibitemShut {NoStop}%
\bibitem [{\citenamefont {Ernst}\ and\ \citenamefont {Hauser}(2021)}]{ernst2021metal}%
  \BibitemOpen
  \bibfield  {author} {\bibinfo {author} {\bibfnamefont {W.~E.}\ \bibnamefont {Ernst}}\ and\ \bibinfo {author} {\bibfnamefont {A.~W.}\ \bibnamefont {Hauser}},\ }\bibfield  {title} {\bibinfo {title} {Metal clusters synthesized in helium droplets: structure and dynamics from experiment and theory},\ }\href@noop {} {\bibfield  {journal} {\bibinfo  {journal} {Phys. Chem. Chem. Phys.}\ }\textbf {\bibinfo {volume} {23}},\ \bibinfo {pages} {7553} (\bibinfo {year} {2021})}\BibitemShut {NoStop}%
\bibitem [{\citenamefont {Movre}\ \emph {et~al.}(2000)\citenamefont {Movre}, \citenamefont {Thiel},\ and\ \citenamefont {Meyer}}]{movre:2000}%
  \BibitemOpen
  \bibfield  {author} {\bibinfo {author} {\bibfnamefont {M.}~\bibnamefont {Movre}}, \bibinfo {author} {\bibfnamefont {L.}~\bibnamefont {Thiel}},\ and\ \bibinfo {author} {\bibfnamefont {W.}~\bibnamefont {Meyer}},\ }\bibfield  {title} {\bibinfo {title} {Theoretical investigation of the autoionization process in molecular collision complexes: {H}e*(2$^3$s)+ {L}i(2$^2$s) → {H}e + {L}i$^+$ + e$^-$},\ }\href@noop {} {\bibfield  {journal} {\bibinfo  {journal} {J. Chem. Phys.}\ }\textbf {\bibinfo {volume} {113}},\ \bibinfo {pages} {1484} (\bibinfo {year} {2000})}\BibitemShut {NoStop}%
\bibitem [{\citenamefont {McKinsey}\ \emph {et~al.}(1999)\citenamefont {McKinsey}, \citenamefont {Brome}, \citenamefont {Butterworth}, \citenamefont {Dzhosyuk}, \citenamefont {Huffman}, \citenamefont {Mattoni}, \citenamefont {Doyle}, \citenamefont {Golub},\ and\ \citenamefont {Habicht}}]{McKinsey:1999}%
  \BibitemOpen
  \bibfield  {author} {\bibinfo {author} {\bibfnamefont {D.~N.}\ \bibnamefont {McKinsey}}, \bibinfo {author} {\bibfnamefont {C.~R.}\ \bibnamefont {Brome}}, \bibinfo {author} {\bibfnamefont {J.~S.}\ \bibnamefont {Butterworth}}, \bibinfo {author} {\bibfnamefont {S.~N.}\ \bibnamefont {Dzhosyuk}}, \bibinfo {author} {\bibfnamefont {P.~R.}\ \bibnamefont {Huffman}}, \bibinfo {author} {\bibfnamefont {C.~E.~H.}\ \bibnamefont {Mattoni}}, \bibinfo {author} {\bibfnamefont {J.~M.}\ \bibnamefont {Doyle}}, \bibinfo {author} {\bibfnamefont {R.}~\bibnamefont {Golub}},\ and\ \bibinfo {author} {\bibfnamefont {K.}~\bibnamefont {Habicht}},\ }\bibfield  {title} {\bibinfo {title} {Radiative decay of the metastable ${\mathrm{he}}_{2}(a{}^{3}{\ensuremath{\Sigma}}_{u}^{+})$ molecule in liquid helium},\ }\href@noop {} {\bibfield  {journal} {\bibinfo  {journal} {Phys. Rev. A}\ }\textbf {\bibinfo {volume} {59}},\ \bibinfo {pages} {200} (\bibinfo {year} {1999})}\BibitemShut {NoStop}%
\bibitem [{\citenamefont {Keto}\ \emph {et~al.}(1974)\citenamefont {Keto}, \citenamefont {Soley}, \citenamefont {Stockton},\ and\ \citenamefont {Fitzsimmons}}]{Keto:1974}%
  \BibitemOpen
  \bibfield  {author} {\bibinfo {author} {\bibfnamefont {J.~W.}\ \bibnamefont {Keto}}, \bibinfo {author} {\bibfnamefont {F.~J.}\ \bibnamefont {Soley}}, \bibinfo {author} {\bibfnamefont {M.}~\bibnamefont {Stockton}},\ and\ \bibinfo {author} {\bibfnamefont {W.~A.}\ \bibnamefont {Fitzsimmons}},\ }\bibfield  {title} {\bibinfo {title} {Dynamic properties of neutral excitations produced in electron-bombarded superfluid helium. ii. afterglow fluorescence of excited helium molecules},\ }\href@noop {} {\bibfield  {journal} {\bibinfo  {journal} {Phys. Rev. A}\ }\textbf {\bibinfo {volume} {10}},\ \bibinfo {pages} {887} (\bibinfo {year} {1974})}\BibitemShut {NoStop}%
\bibitem [{\citenamefont {Carter}\ \emph {et~al.}(2017)\citenamefont {Carter}, \citenamefont {Hertel}, \citenamefont {Rooks}, \citenamefont {McClintock}, \citenamefont {McKinsey},\ and\ \citenamefont {Prober}}]{carter:2017}%
  \BibitemOpen
  \bibfield  {author} {\bibinfo {author} {\bibfnamefont {F.}~\bibnamefont {Carter}}, \bibinfo {author} {\bibfnamefont {S.}~\bibnamefont {Hertel}}, \bibinfo {author} {\bibfnamefont {M.}~\bibnamefont {Rooks}}, \bibinfo {author} {\bibfnamefont {P.}~\bibnamefont {McClintock}}, \bibinfo {author} {\bibfnamefont {D.}~\bibnamefont {McKinsey}},\ and\ \bibinfo {author} {\bibfnamefont {D.}~\bibnamefont {Prober}},\ }\bibfield  {title} {\bibinfo {title} {Calorimetric observation of single {H}e$_2^*$ excimers in a 100-mk {H}e bath},\ }\href@noop {} {\bibfield  {journal} {\bibinfo  {journal} {J. Low Temp. Phys.}\ }\textbf {\bibinfo {volume} {186}},\ \bibinfo {pages} {183} (\bibinfo {year} {2017})}\BibitemShut {NoStop}%
\bibitem [{\citenamefont {{\v{C}}erm{\'a}k}(1966)}]{vcermak1966individual}%
  \BibitemOpen
  \bibfield  {author} {\bibinfo {author} {\bibfnamefont {V.}~\bibnamefont {{\v{C}}erm{\'a}k}},\ }\bibfield  {title} {\bibinfo {title} {Individual efficiency curves for the excitation of 23s and 21s states of helium by electron impact},\ }\href@noop {} {\bibfield  {journal} {\bibinfo  {journal} {J. Chem. Phys.}\ }\textbf {\bibinfo {volume} {44}},\ \bibinfo {pages} {3774} (\bibinfo {year} {1966})}\BibitemShut {NoStop}%
\bibitem [{\citenamefont {Dugan}\ \emph {et~al.}(1967)\citenamefont {Dugan}, \citenamefont {Richards},\ and\ \citenamefont {Muschlitz~Jr}}]{dugan1967excitation}%
  \BibitemOpen
  \bibfield  {author} {\bibinfo {author} {\bibfnamefont {J.}~\bibnamefont {Dugan}}, \bibinfo {author} {\bibfnamefont {H.}~\bibnamefont {Richards}},\ and\ \bibinfo {author} {\bibfnamefont {E.~E.}\ \bibnamefont {Muschlitz~Jr}},\ }\bibfield  {title} {\bibinfo {title} {Excitation of the metastable states of helium by electron impact},\ }\href@noop {} {\bibfield  {journal} {\bibinfo  {journal} {J. Chem. Phys.}\ }\textbf {\bibinfo {volume} {46}},\ \bibinfo {pages} {346} (\bibinfo {year} {1967})}\BibitemShut {NoStop}%
\bibitem [{\citenamefont {Ralchenko}\ \emph {et~al.}(2008)\citenamefont {Ralchenko}, \citenamefont {Janev}, \citenamefont {Kato}, \citenamefont {Fursa}, \citenamefont {Bray},\ and\ \citenamefont {de~Heer}}]{Ralchenko:2008}%
  \BibitemOpen
  \bibfield  {author} {\bibinfo {author} {\bibfnamefont {Y.}~\bibnamefont {Ralchenko}}, \bibinfo {author} {\bibfnamefont {R.}~\bibnamefont {Janev}}, \bibinfo {author} {\bibfnamefont {T.}~\bibnamefont {Kato}}, \bibinfo {author} {\bibfnamefont {D.}~\bibnamefont {Fursa}}, \bibinfo {author} {\bibfnamefont {I.}~\bibnamefont {Bray}},\ and\ \bibinfo {author} {\bibfnamefont {F.}~\bibnamefont {de~Heer}},\ }\bibfield  {title} {\bibinfo {title} {Electron-impact excitation and ionization cross sections for ground state and excited helium atoms},\ }\href@noop {} {\bibfield  {journal} {\bibinfo  {journal} {At. Data Nucl. Data Tables}\ }\textbf {\bibinfo {volume} {94}},\ \bibinfo {pages} {603} (\bibinfo {year} {2008})}\BibitemShut {NoStop}%
\bibitem [{\citenamefont {Miteva}\ \emph {et~al.}(2017)\citenamefont {Miteva}, \citenamefont {Kazandjian},\ and\ \citenamefont {Sisourat}}]{miteva:2017}%
  \BibitemOpen
  \bibfield  {author} {\bibinfo {author} {\bibfnamefont {T.}~\bibnamefont {Miteva}}, \bibinfo {author} {\bibfnamefont {S.}~\bibnamefont {Kazandjian}},\ and\ \bibinfo {author} {\bibfnamefont {N.}~\bibnamefont {Sisourat}},\ }\bibfield  {title} {\bibinfo {title} {On the computations of decay widths of fano resonances},\ }\href@noop {} {\bibfield  {journal} {\bibinfo  {journal} {Chem. Phys.}\ }\textbf {\bibinfo {volume} {482}},\ \bibinfo {pages} {208} (\bibinfo {year} {2017})}\BibitemShut {NoStop}%
\bibitem [{\citenamefont {Brauer}\ \emph {et~al.}(2013)\citenamefont {Brauer}, \citenamefont {Smolarek}, \citenamefont {Loginov}, \citenamefont {Mateo}, \citenamefont {Hernando}, \citenamefont {Pi}, \citenamefont {Barranco}, \citenamefont {Buma},\ and\ \citenamefont {Drabbels}}]{Brauer:2013}%
  \BibitemOpen
  \bibfield  {author} {\bibinfo {author} {\bibfnamefont {N.~B.}\ \bibnamefont {Brauer}}, \bibinfo {author} {\bibfnamefont {S.}~\bibnamefont {Smolarek}}, \bibinfo {author} {\bibfnamefont {E.}~\bibnamefont {Loginov}}, \bibinfo {author} {\bibfnamefont {D.}~\bibnamefont {Mateo}}, \bibinfo {author} {\bibfnamefont {A.}~\bibnamefont {Hernando}}, \bibinfo {author} {\bibfnamefont {M.}~\bibnamefont {Pi}}, \bibinfo {author} {\bibfnamefont {M.}~\bibnamefont {Barranco}}, \bibinfo {author} {\bibfnamefont {W.~J.}\ \bibnamefont {Buma}},\ and\ \bibinfo {author} {\bibfnamefont {M.}~\bibnamefont {Drabbels}},\ }\bibfield  {title} {\bibinfo {title} {Critical landau velocity in helium nanodroplets},\ }\href@noop {} {\bibfield  {journal} {\bibinfo  {journal} {Phys. Rev. Lett.}\ }\textbf {\bibinfo {volume} {111}},\ \bibinfo {pages} {153002} (\bibinfo {year} {2013})}\BibitemShut {NoStop}%
\end{thebibliography}%

\end{document}


\title{Supplementary Material\linebreak
Interatomic Coulombic decay in large lithium-doped helium nanodroplets induced by photoelectron impact excitation}

\author{L. Ben Ltaief \textit{et al.}}

\maketitle

\renewcommand\refname{SM References}
\renewcommand{\figurename}{SM Fig.}

\date{}
\section{Electron spectra}
\begin{figure}[!h]
	\center
    \includegraphics[width=0.35\columnwidth]{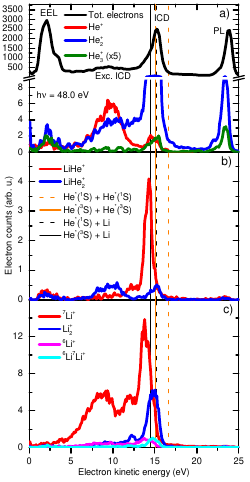}\caption{\label{fig5} Electron spectra measured for Li doped large He nanodroplets irradiated with photons of energy $h\nu$ = 48.0~eV. a) Electron spectra measured in coincidence with He$^+$ [red line], He$_2^+$ [blue line] and He$_3^+$ [green line]. b) Electron spectra measured in coincidence with LiHe$^+$ [red line] and LiHe$_2^+$ [blue line]. c) Electron spectra measured in coincidence with $^7$Li$^+$ [red line], $^7$Li$_2^+$ [blue line], $^6$Li$^+$ [pink line] and $^6$Li$^7$Li$^+$ [cyan line]. The spectrum plotted in black in a) is for all electrons. All spectra are normalized to the photon flux and to the $h\nu$-dependent absorption cross section of He.}
\end{figure}
To see whether higher-lying excited states of He$^*$ (1s$n$p, $n>2$) are involved in the ICD reactions described by Eqs. (1) and (2) in the main text, electron-ion coincidence measurements are repeated for Li-doped He nanodroplets irradiated with XUV photon of energy $h\nu = 48.0$~eV. SM Fig. 1 shows total and coincidence electron spectra. 
The spectra are nearly identical as those shown in Fig.~4 of the main text, indicating that  He$^*$'s formed by the electron-impact excitation and electron-ion recombination fully relax mostly to the lowest metastable He$^*$ in 1s2s\,$^3$S and to He$_2^*$ in a$^3\Sigma_u^+$ states prior to ICD.

 \section{Electron and ion yields}
 To support Fig. 5 in the main text, the yields of Li dopant ICD electrons vs. He droplet ICD electrons were also measured across the electron-impact excitation threshold in the photon energy range $h\nu = 44$-$48.4$~eV for Li doped large He nanodroplets of droplet size $R= 75$~nm, see SM Fig. 2. To get further details about the efficiency of ICD for different droplet sizes across the electron-impact excitation threshold, the same photon energy-dependent ion and ICD electron yield measurements were also repeated for a droplet size $R= 35$~nm as shown in SM Fig. 3 and SM Fig. 4, respectively. All these ions and electrons yields show the same behaviour as those measured for a droplet size $R= 75$~nm; except that they appeared a bit more suppressed above their maxima as for small He droplets one would expect that a significant number of He atoms excited in their higher states tend to be desorbed from the He droplet surface before they relax to lower states and then decay by ICD. Furthermore, the ions yields appeared more enhanced below their onsets (see SM Fig. 3) as the ejection mechanism of He photions and Li ions (being created by electron transfer) is more efficient for small He droplets as compared to large He droplets where most of the ions tend to sink inside instead.
\begin{figure}[!h]
	\center
    \includegraphics[width=0.5\columnwidth]{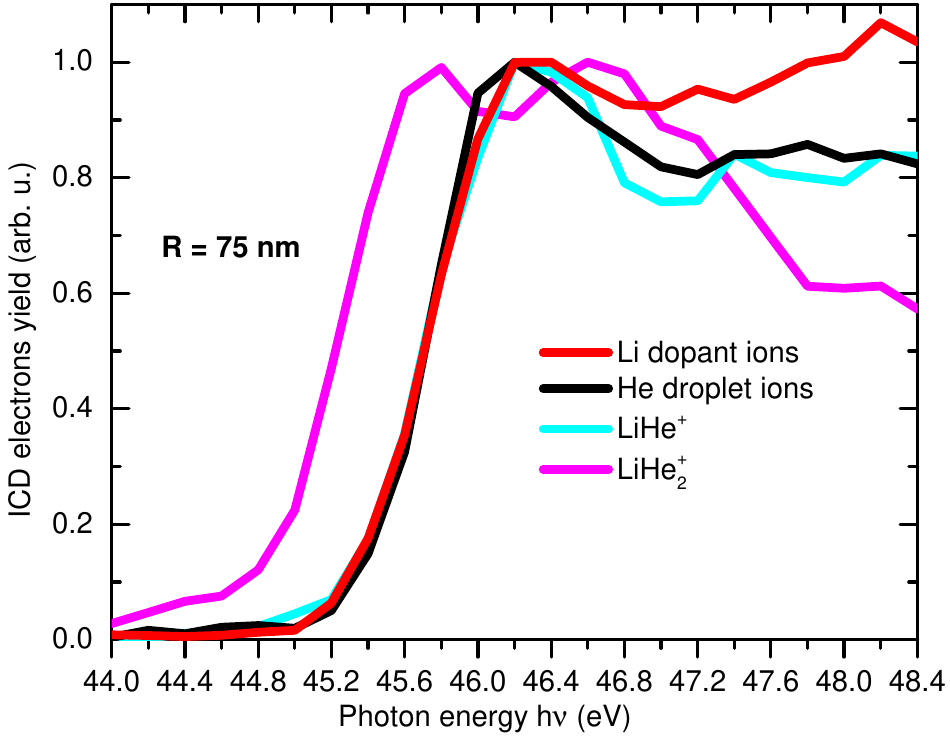}\caption{\label{fig:hv-scanelectrons} Photon energy-dependent yields of ICD electrons measured in coincidence with He droplet ions, Li dopant ions, LiHe$^+$ and LiHe$_2^+$ for Li doped large He nanodroplets ($R= 75$~nm). The red curve is the sum of yields of electrons measured in coincidence with $^7$Li$+$, $^6$Li$+$, Li$_2^+$ and $^6$Li$^7$Li$^+$ ions. The black curve is the sum of yields of electrons measured in coincidence with He$^+$, He$_2^+$ and He$_3^+$. The yield of electrons measured in coincidence with LiHe$^+$ and LiHe$_2^+$ is shown in the curves plotted in cyan and magenta, respectively. The gray curve shows the yield of all ICD electrons. All the data shown here are normalized to the photon flux and to the $h\nu$-dependent absorption cross section of He.}
\end{figure}
 \begin{figure}[!h]
	\center
    \includegraphics[width=0.5\columnwidth]{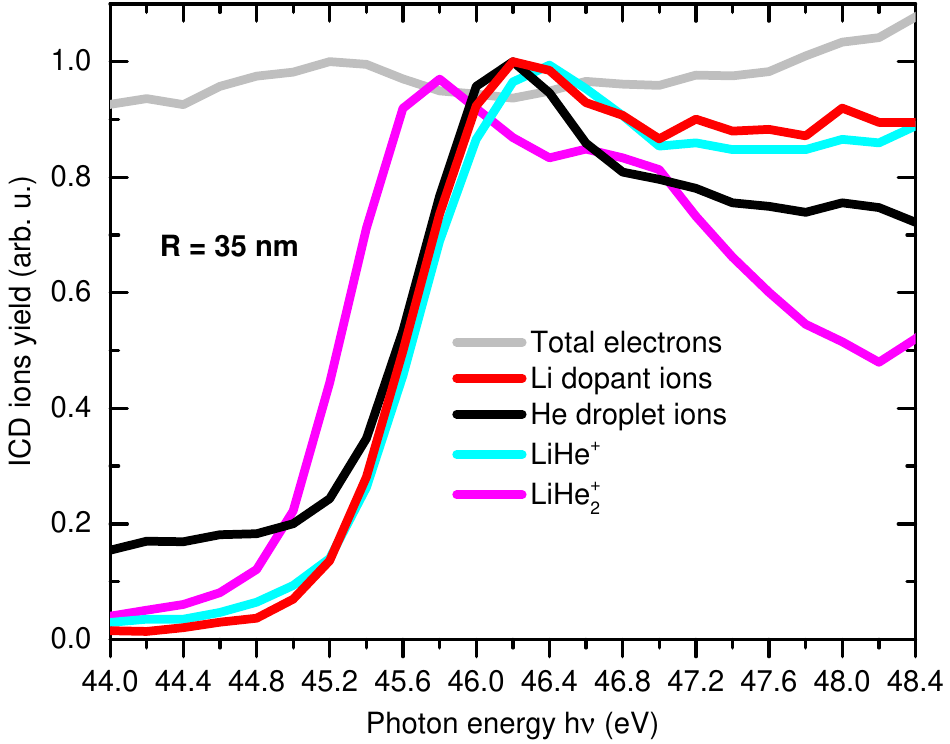}\caption{\label{Supp_Fig_3} Photon energy-dependent ion yields measured for Li doped large He nanodroplet of droplet size $R= 35$~nm. Red line: yield of all ions containing Li. Black line: sum of He$_n^+$ ion yields for $n=1$-3. Cyan line: yield of [LiHe]$^+$ ions. Magenta line: yield of [LiHe$_2$]$^+$ ions.}
\end{figure}
\begin{figure}[!h]
	\center
    \includegraphics[width=0.5\columnwidth]{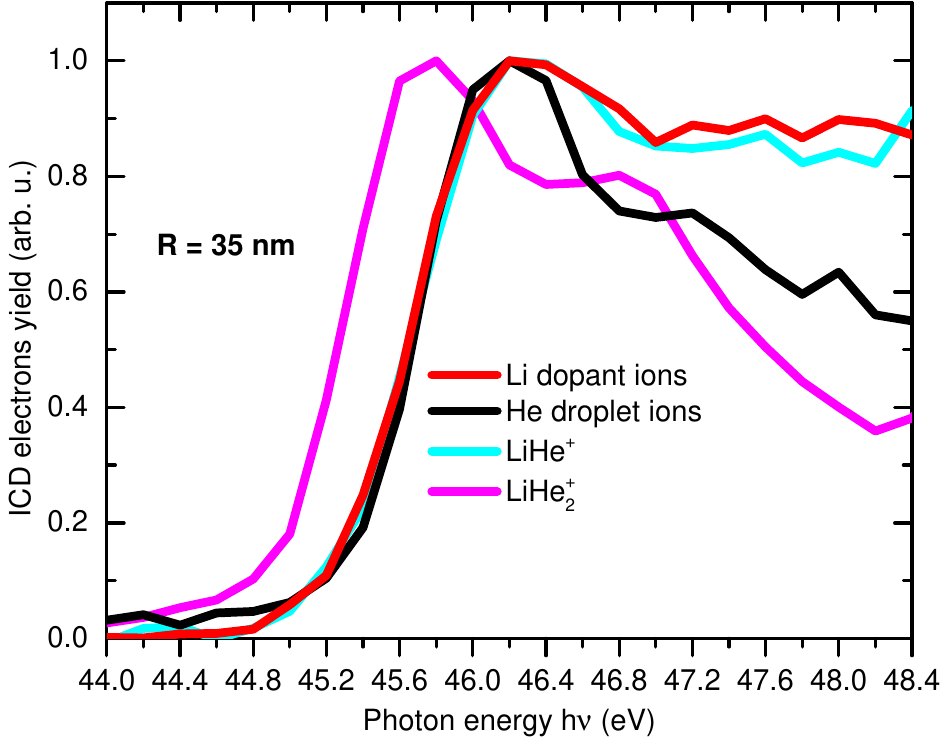}\caption{\label{Supp_Fig_1} Photon energy-dependent yields of ICD electrons measured in coincidence with He droplet ions, Li dopant ions, LiHe$^+$ and LiHe$_2^+$ for Li doped large He nanodroplets with droplet size of $R= 35$~nm. The red curve is the sum of yields of electrons measured in coincidence with $^7$Li$+$, $^6$Li$+$, Li$_2^+$ and $^6$Li$^7$Li$^+$ ions. The black curve is the sum of yields of electrons measured in coincidence with He$^+$, He$_2^+$ and He$_3^+$. The yield of electrons measured in coincidence with LiHe$^+$ and LiHe$_2^+$ is shown in the curves plotted in cyan and magenta, respectively. The gray curve shows the yield of all ICD electrons. All the data shown here are normalized to the photon flux and to the $h\nu$-dependent absorption cross section of He.}
\end{figure}

\section{Photon energy-dependent total electrons spectra}
The total electron yield shown in the main text in Fig. 5 as a function of the photon energy in the range $h\nu=$~44.0-48.4~eV is obtained by integrating over all the spectra measured for all the electrons emitted from Li-doped large He nanodroplets ($R= 75$~nm). Typical total electron spectra measured in a wider photon energy range $h\nu=43.6$-49.8~eV and in steps of 0.2~eV are shown in SM Fig. 5. Clearly, by varying the photon energy from $h\nu=$~43.6~eV up to higher photon energies the He photoline (PL) becomes less pronounced at $h\nu\geq$~44.0~eV, whereas the ICD electron and the electron energy loss (EEL) features begun to merge at $h\nu=$~45.0~eV and become more and more visible at $h\nu>$~45.0~eV.

\begin{figure}[!h]
	\center
\includegraphics[width=0.5\columnwidth]{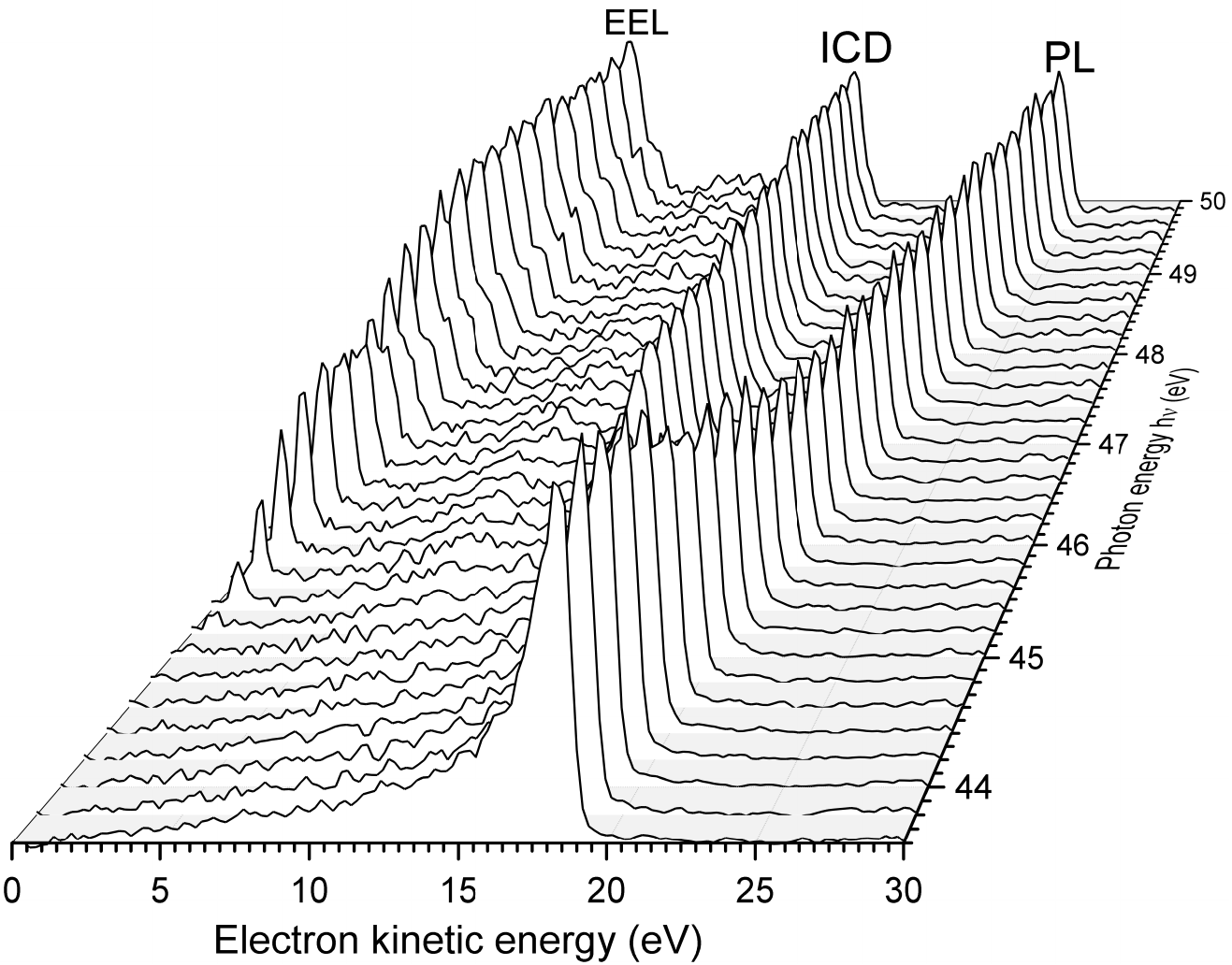}\caption{\label{Supp_Fig_5}
Photon energy-dependent spectra of all electrons measured for Li-doped large He nanodroplets ($R= 75$~nm). The positions of the He photoline (PL) and the electron energy loss line change as a function of photon energy, whereas the position of the ICD peak remains constant at 15.0~eV. All the spectra shown here are normalized to the photon flux and the photon energy-dependent photoionization cross section of He.}
\end{figure}

\section{Contribution of He $^1$S states in ICD}
According to \cite{Ralchenko:2008,register:1980}, at kinetic energies of incident electrons $>22$~eV one would expect that more He$^*$ atoms are excited to their singlet states rather than in their triplet states as the cross section of electron impact excitation leading to formation of singlet states becomes significantly higher at these electron energies. However, our spectra measured at $h\nu = 46.2$~eV (see Fig. 3 and 4 b) in the main text) show that only triplet states contribute to ICD. To check whether at photon energies higher than $h\nu = 46.2$~eV there is a contribution of singlet states in ICD, we recorded ICD electron spectra in coincidence with $^7$Li$^+$ at $h\nu = 46.4$, 48.0, 49.6, 51.2 and 52.8~eV. Starting from $h\nu = 48.0$~eV, all recorded spectra clearly show a weak shoulder around 15.2 eV, see SM Fig. 5. We attribute this shoulder to ICD involving one He$^*$ in the $^1$S state and one Li atom. The reason for the small contribution of ICD involving He$^*$($^1$S) even at these higher photon energies is not fully resolved. The ICD rate seems to be higher for the $^3$S state by a factor of 5; possibly the shorter lifetime of the He$^*$($^1$S) of a few ns due to spontaneous decay by emission of fluorescence is a limiting factor. 

\begin{figure}[!h]
	\center
\includegraphics[width=0.6\columnwidth]{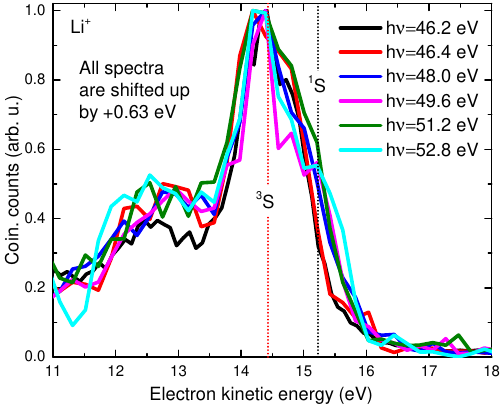}\caption{\label{Supp_Fig_5}Spectra of ICD electrons measured for Li doped large He nanodroplets ($R= 75$~nm) and in coincidence with $^7$Li$+$ ions at different photon energies $h\nu$= 46.2~eV (black spectrum), 46.4~eV (red spectrum), 48.0~eV (blue spectrum), 49.6~eV (magenta spectrum), 51.2~eV (green spectrum)and 52.8~eV (cyan spectrum). The dashed vertical straight lines in red and black indicate characteristic electron energies for ICD out of $^3$S and $^1$S states, respectively.}
\end{figure}

\bibliography{HeDropletPI_Bib}